\documentclass[aps,prb,reprint, amsmath, amssymb]{revtex4-1}

\usepackage{bm}
\usepackage{booktabs}
\usepackage[retainorgcmds]{IEEEtrantools}
\usepackage{graphicx}
\usepackage{mathrsfs}
\usepackage{amsmath}
\usepackage{amssymb}
\usepackage{color}
\usepackage{amsfonts}

\newcommand{\ud}{\,\mathrm{d}}

\newcommand{\cm}{\chi_\text{max}''}

\begin{document}

\title{AC susceptibility as a tool to probe the dipolar interaction in magnetic nanoparticles}
\date{\today}
\author{Gabriel T. Landi}
\email{gtlandi@gmail.com}
\author{Fabiana R. Arantes}
\affiliation{Universidade Federal do ABC,  09210-580 Santo Andr\'e, Brazil}
\author{Daniel R. Cornejo}
\affiliation{Instituto de F\'isica da Universidade de S\~ao Paulo, S\~ao Paulo 05508-090, Brazil}
\author{Andris F. Bakuzis}
\affiliation{Instituto de F\'isica, Universidade Federal de Goi\'as, 74001-970, Goi\^ania-GO, Brazil}
\author{Irene Andreu}
\author{Eva Natividad}
\affiliation{Instituto de Ciencia de Materiales de Arag\'on (ICMA), CSIC-Universidad de Zaragoza, Zaragoza  50018, Spain}

\begin{abstract}

The dipolar interaction is known to substantially affect the properties of magnetic nanoparticles.
This is particularly important when the particles are kept in a fluid suspension or packed inside nano-carriers. 
In addition to its usual long-range nature, in these cases the dipolar interaction may also  induce the formation of clusters of particles, thereby  strongly modifying   their magnetic anisotropies. 
In this paper we show how AC susceptibility may be used to obtain important information regarding the influence of the dipolar interaction in a sample. 
We develop a model which includes both aspects of the dipolar interaction and may be fitted directly to the susceptibility data. 
The usual long-range nature of the interaction is implemented using a mean-field solution, whereas the particle-particle aggregation is modeled using a distribution of anisotropy constants. 
The model is then applied to two samples studied at different concentrations. 
One consists of spherical magnetite nanoparticles dispersed in oil and the other of cubic magnetite nanoparticles embedded on PLGA nanospheres. 
We also introduce a simple technique to access the importance of the dipolar interaction in a given sample, based on the height of the AC susceptibility peaks for different driving frequencies. Our results help illustrate the important effect that the dipolar interaction has in most nanoparticle samples.

\end{abstract}
\maketitle{}

%
%
%
%

\section{\label{sec:int}Introduction}

Magnetic nanoparticles (MNPs) have been an active topic of research for over half a century. 
Initially, much of this interest was related to the magnetic recording industry, but in the past few decades there has been a shift toward biologically inclined applications.\cite{Krishnan2010}
Examples include the use of MNPs for drug delivery,\cite{Dobson2006,Nobuto2004,Babincova2002}
stem cell labeling,\cite{Lu2007,Bulte2001}
water treatments,\cite{Yavuz2006}
contrast agents for nuclear magnetic resonance\cite{Larsen2008}
and magnetic hyperthermia.
\cite{Gilchrist1957,Hilger2005,Pankhurst2009,Lee2011c,Branquinho2013,Andreu2014,Andreu2015,Rodrigues2013,Dennis2009a,Maier-Hauff2011a}
The latter, in particular, is a cancer treatment technique that has already entered clinical trials\cite{Maier-Hauff2011a} and is now considered the most promising application of MNPs.
Great progress has also been made in our theoretical understanding of MNPs, particularly through Brown's Fokker-Planck equation,\cite{Brown1963,*Brown1979a,Coffey2004} 
which has allowed us to make valuable predictions about many  dynamic properties,\cite{Coffey1993,*Coffey2013,
Shliomis1993,
Garcia-Palacios1998,*Garcia-Palacios2000,
Kalmykov1999a,*Kalmykov2010,
Usov2009,*Usov2009a,
Dejardin2010,
Poperechny2010,
Ouari2013,
Landi2011c,*Landi2012b,*Landi2012e}
therefore providing a microscopic platform to describe  hyperthermia experiments.\cite{Carrey2011,Verde2012,*Verde2012a,*Landi2012f}

Most of our theoretical understanding about MNPs concerns non-interacting samples. 
However, due to their large magnetic moments, MNPs are also strongly influenced by the dipolar interaction.
\cite{
Shtrikman1981,
Chantrell1983,
Jonsson1995,
Jonsson2001,
Dormann1996,*Dormann1999,
Andersson1997,
Kechrakos1998a,Kechrakos2002,
Masunaga2009,*Masunaga2011b,
Nadeem2011,
SungLee2011,
Dejardin2011a,
Felderhof2003,
Jonsson2004,
Dennis2008,
Urtizberea2010,
Haase2012,
Landi2013,*Landi2014,
Ruta2015,
Lyberatos1993a,
Zhang1999,*Zhang2000,
Denisov2003,
Titov2005,
Mao2008,
Serantes2010a,
Mehdaoui2013a,*Tan2014}
Indeed, recent papers\cite{Landi2013,*Landi2014,Mehdaoui2013a,*Tan2014,Ruta2015,Branquinho2013,Andreu2014,Andreu2015} have shown that the dipolar interaction has a strong influence in magnetic hyperthermia treatments. 
This means that the heating properties of particles diluted in a fluid will be very different from those of particles packed inside cells or  nano-carriers, such as magnetoliposomes.\cite{DeCuyper1988,DiCorato2014a,Andreu2015,Cintra2009}
Hence, when tailoring a sample for a specific treatment, one must also take into account the spatial arrangement  of the nanoparticles.
Recently,  several theoretical models \cite{Shtrikman1981,Branquinho2013,Landi2013,*Landi2014,Ruta2015} and  simulations methods\cite{Mehdaoui2013a,*Tan2014} have been developed to deal with the dipolar interaction and aid in the design of samples for specific treatments.

However, in certain samples the dipolar interaction may also be responsible for an indirect effect, which is seldom taken into account when developing theoretical models. 
Namely, it induces the aggregation of  clusters of MNPs  (sometimes observed in the form of elongated chains\cite{Bakuzis2013,*Castro2008,*Eloi2010,Branquinho2013}).
The strong interaction between particles within a cluster cause them to rotate in order to align their easy axes, therefore modifying (usually increasing) substantially their effective magnetic anisotropy.\cite{Jacobs1955,Branquinho2013}
This effect exists on top of the usual dipolar interaction, but may lead to quite distinct consequences.
It is also extremely common for samples used in hyperthermia.

The modifications in the effective magnetic anisotropy of a given MNP will depend sensibly on the  size and shape of the aggregate that it resides in, and also on the position of that MNP within the cluster. 
Thus, in any given sample, one may expect a  broad and highly complex distribution of anisotropy constants, in addition to the  distribution of volumes. 
Of course,  a distribution of anisotropies will exist even for non-interacting samples due to the fluctuations in the crystallinity, shape and surface roughness of the particles. 
However, due to  recent  improvements in sample preparation methods, these intrinsic effects have been substantially minimized and therefore should be negligible in comparison with the fluctuations  brought about by  particle-particle aggregation.


Experimentally accessing and quantifying the degree of aggregation, however, is by no means trivial.
This problem has generated much interest lately, with recent proposals involving the use of Lorentz microscopy\cite{Campanini2015} and small angle X ray scattering.\cite{Coral2016}
The purpose of this paper is to show that AC susceptibility measurements, a technique which is easily accessible experimentally, can also yield important information concerning the state of aggregation in a  sample.

Traditionally, AC susceptibility curves are analyzed by looking at the temperature $T_\text{max}$ where the imaginary part $\chi''$ is a maximum. 
An analysis of $T_\text{max}$ as a function of the frequency $f$ of the AC field is then used to extract information about the energy barrier distribution in the sample. 
This analysis, referred to as an Arrhenius plot, clearly underuses the data since from each $\chi''$~vs.~$T$ dataset, just a single point is taken (the maximum). 
It also does not explicitly include the effects of the particle size distribution. 
A more satisfactory approach is that of Jonsson \emph{et. al.},\cite{Jonsson1997} which developed a model  that can be fitted to the entire dataset, taking into account the size distribution.

In this paper we discuss how to expand on the model of Ref.~\onlinecite{Jonsson1997} to include  both aspects of the dipolar interaction.
First, the long-range effect is implemented using three known models:  
the Vogel-Fulcher approximation\cite{Shtrikman1981}, a mean-field model developed recently by one of the authors\cite{Landi2013,*Landi2014}  and the Dormann-Bessais-Fiorani (DBF) model.\cite{Dormann1999}
Second, the effect of particle aggregation is taken into account by introducing a distribution of anisotropy constants.  
We show that it is possible  to exploit the morphological information extracted from TEM to isolate in the AC susceptibility analysis the contribution of the anisotropy distribution induced by the  the aggregation process. 
As a result, we are able to extract information which reflects the different levels of particle aggregation within the sample.

We also introduce a new very simple tool to access the qualitative importance of the dipolar interaction in a given sample. 
It is based on analyzing the maximum height  $\cm$  of the $\chi''$~vs.~$T$ curves  as a function of the frequency $f$. 
When the dipolar interaction is negligible in a sample,  $\cm$ never increases with $f$. 
Conversely, we show that the presence of a dipolar interaction causes $\cm$ to \emph{increase} with $f$. 
Hence, this serves as a signature of the dipolar interaction. 
Through a simple visual analysis  of the imaginary AC susceptibility curves it is possible to see if the dipolar interaction is important in that given sample or not.
We believe that this simple test may substantially help researchers in accessing the extent of the dipolar interaction in a sample.

We apply this model to two samples, a commercial magnetite-based ferrofluid dispersed in oil, and a sample containing PLGA nanospheres packed with cubic magnetite  nanoparticles. 
For the commercial ferrofluid, we find that the distribution of energy barriers is bimodal, with the vast majority of particles living in large aggregates and a small fraction still in free suspension in the fluid. 
For the PLGA nanospheres, due to the tightly packed nature of the nanoparticles,  we observe a much more complex energy barrier distribution containing at least 3 distinct aggregate configurations. 
Given the importance of AC susceptibility for many applications of MNPs,\cite{Herrera2010,Nutting2006,Fannin1986,*Fannin2003}  we believe that the present paper may be of value to researchers working with MNPs and biomagnetism.

%
%
%
%

\section{\label{sec:theory}Theory}

%
%
%
%

\subsection{\label{ssec:monoAC}AC susceptibility for ideal monodisperse samples}

Consider a sample of single-domain magnetic nanoparticles, all having a volume $V$ and uniaxial anisotropy constant $K$. 
The relaxation time of the particles at a temperature $T$ is  given approximately by the N\'eel formula:\cite{Neel1949}$^{,}$\footnote{\label{note1}
The relaxation time may be computed exactly for single-domain particles using the Fokker-Planck equation, but the solution is expressed in terms of complex hypergeometric functions.\cite{Coffey1994}
An approximate formula which is more precise than Eq.~(\ref{tau1})  was first given by Brown\cite{Brown1963,*Brown1979a} and reads
$\tau = \frac{\tau_0}{2} \sqrt{\frac{\pi}{\sigma}} e^\sigma$. 
However, both this formula and Eq.~(\ref{tau1}) suffer from the deficiency that they are not zero when $\sigma\to 0$, which should be true since in this case there is no energy barrier to surmount. In fact, they are good approximations to the real relaxation time  only for $\sigma > 2$. A formula which is extremely precise, and valid for all values of $\sigma$, is\cite{Coffey2013}
\[
\tau = \tau_0 \frac{e^{\sigma }-1}{\frac{2 \sigma ^{3/2}}{\sqrt{\pi } (\sigma +1)}+2^{-\sigma }}
\]
}
\begin{equation}\label{tau1}
\tau \simeq \tau_0 e^{\sigma}
\end{equation}
where $\tau_0 \sim 10^{-9}$ s and 
\begin{equation}\label{sigma}
\sigma = \frac{\theta}{T} = \frac{KV}{k_B T}
\end{equation}
The quantity $\theta = KV/k_B$, which will be used throughout the text,  represents the height of the energy barrier in temperature units.

In AC susceptibility experiments one measures the response of a sample to an alternating magnetic field $H(t) = H_0 \cos\omega t$, of frequency $f = \omega/2\pi$ and very low amplitude $H_0$. 
In this case it is known from  linear response theory  (see Appendix~\ref{app:lin}) that the component of the magnetic moment of the sample in the direction of the exciting field, $\mu(t)$, is described by:
\begin{equation}\label{mu_t}
\mu(t) = H_0 (\chi' \cos\omega t + \chi'' \sin\omega t)
\end{equation}
That is, $\mu(t)$ will try to follow $H(t)$, but will do so with a phase lag. 
In this equation, 
\begin{equation}\label{CR}
\chi' = \chi_0 \frac{1}{1+(\omega \tau)^2} 
\end{equation}
is the real (in-phase) component of the dynamic susceptibility and 
\begin{equation}\label{CI} 
\chi'' =  \chi_0 \frac{\omega\tau}{1+(\omega \tau)^2}
\end{equation}
is the imaginary (out-of-phase) component. Moreover,  $\chi_0$ is the static susceptibility. 

Notice that in Eq.~(\ref{mu_t}) we are describing the response in terms of magnetic moments $\mu$ and not magnetization $M = \mu/V$.
The reason for this will be clarified   in Sec.~\ref{ssec:poli}. 
Due to this choice, the  static susceptibility $\chi_0$ for randomly-oriented single-domain particles correspond to the Langevin susceptibility times the particle's volume:
\begin{equation}\label{chi_0}
\chi_0 = \frac{(M_s V)^2}{3 k_B T}
\end{equation}

\begin{figure}[!h]
\centering
\includegraphics[width=0.22\textwidth]{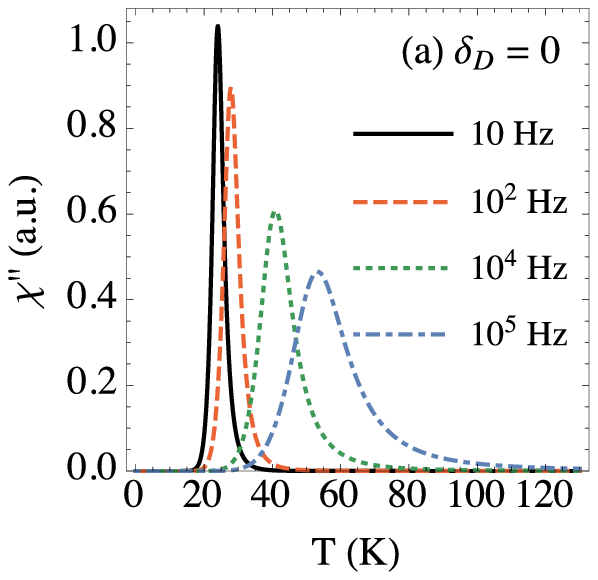}\quad
\includegraphics[width=0.22\textwidth]{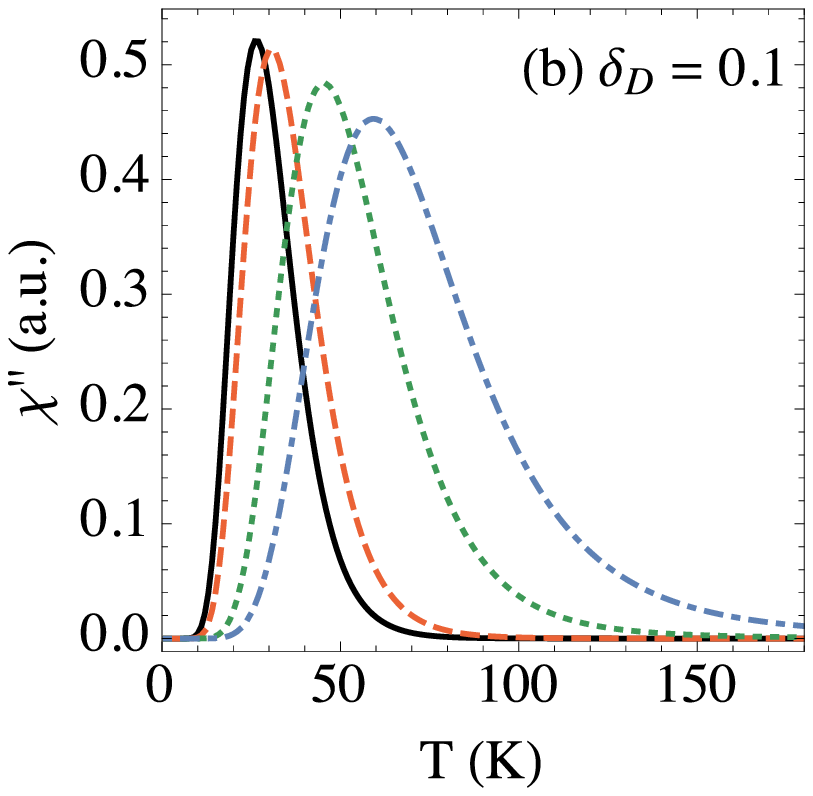}\\
\includegraphics[width=0.22\textwidth]{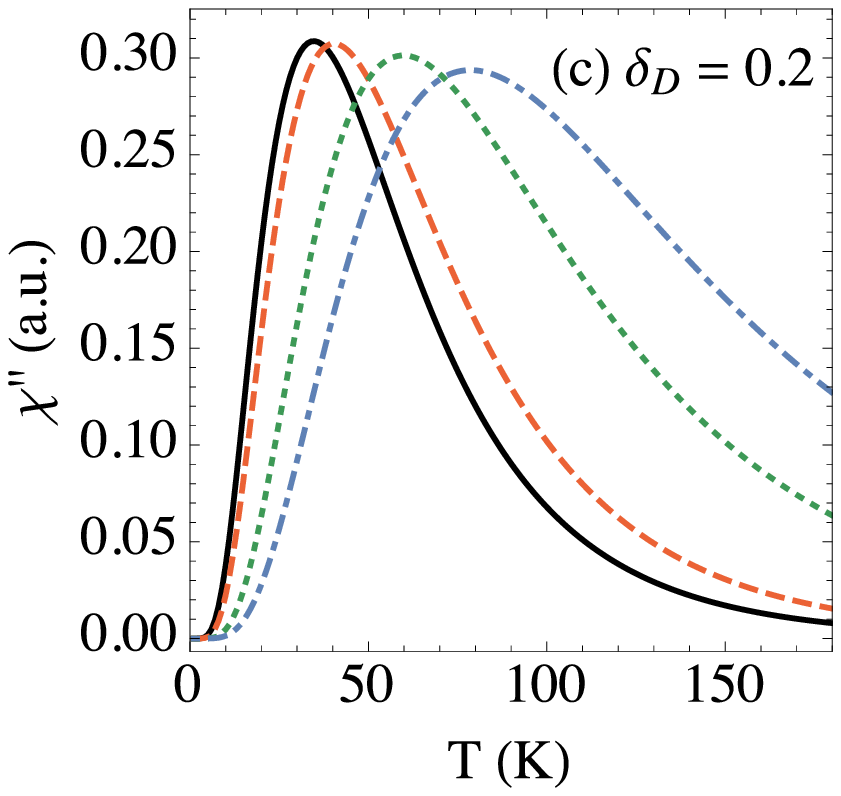}\quad
\includegraphics[width=0.22\textwidth]{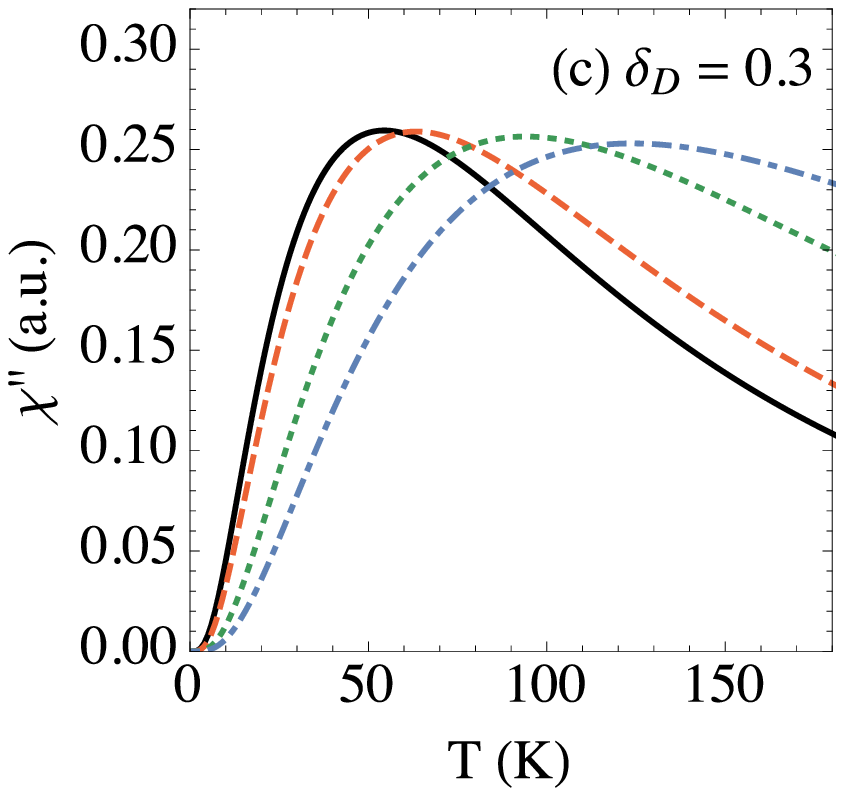}\\
\caption{\label{fig:example}(Color Online)
(a) Example curves of $\chi''$~vs.~$T$ for ideal monodisperse samples computed from Eq.~(\ref{CI}) with $\theta = 400$ K, $\tau_0 = 10^{-9}$ s and different values of $f$. 
(b)-(d) Same, but for polidisperse samples, computed from the numerical solution of the integral in Eq.~(\ref{1}), with $\theta_0 = 400$ K and three different values of $\delta_D$.
}
\end{figure}

In Fig.~\ref{fig:example}(a) we show examples of Eq.~(\ref{CI}) for $\theta = 400$ K, $\tau_0 = 10^{-9}$ s and different frequencies $f$. 
As can be seen, these curves do not resemble commonly encountered experimental data [cf. Figs.~\ref{fig:fabi_data} and \ref{fig:irene_data}, or Refs.~\onlinecite{Nadeem2011,Urtizberea2010}]. 
This is due to the fact that we are ignoring the size distribution of the MNPs and the magnetic dipolar interaction.

Experimental curves of $\chi''$ vs.~$T$ are usually analyzed  by looking at the temperature where $\chi''$ is a maximum. 
According to Eq.~(\ref{CI}) this occurs at $\omega\tau = 1$, which implies the equation
\begin{equation}\label{arrhenius}
-\ln(2\pi f) = \ln (\tau_0) + \frac{\theta}{T_\text{max}}
\end{equation}
Hence, a plot of $-\ln(2\pi f)$~vs.~$1/T_\text{max}$ should yield a straight line, from which it is possible to extract $\tau_0$ and $\theta$. 
This is usually referred to as an Arrhenius plot. 

The Arrhenius plot clearly underuses the available information, since from the entire data set only a single point is used (the maximum). 
Moreover, it is also very sensitive to experimental uncertainties. 
A more robust approach, in which a model can be fitted to the entire data set, is that developed by Jonsson \emph{et. al.} \cite{Jonsson1997}, which will be described  in the Sec.~\ref{ssec:poliAC}.
But before we do so, we must first generalize the above results to include the effects of a size distribution. 
%

%
%
%
%

\subsection{\label{ssec:poli}Magnetic moment of samples with a size distribution}

The case of ideal monodisperse samples, where all particles have the exact same volume, is highly improbable in practice. 
For example, in Ref.~\onlinecite{Park2005} a minimum standard deviation of 3\% in the diameter was obtained, even though the synthetic method used 
is known from the literature to produce highly monodisperse MNPs. 
It is therefore essential to include in any analysis the effects of a size distribution. 

If we randomly draw a particle from a sample and measure its diameter $D$, the result will  be a random quantity described by a probability distribution $P(D)$. 
This distribution may be experimentally accessed from  TEM measurements  by directly counting the number of particles with a given diameter (cf.  Figs.~\ref{fig:fabi_size} and \ref{fig:irene_size} below).

It is also customary to model $P(D)$ by a lognormal distribution\footnote{
This assumption is based on empirical evidences and the fact that the lognormal distribution satisfies certain convenient properties. 
Sometimes it is found that a Gaussian distribution also adequately describes the data, but for the purpose of modeling, this is not satisfactory since the Gaussian distribution would allow for negative diameters. 
If a more general formula is necessary, a useful alternative is the Gamma distribution. 
} 
\begin{equation}\label{PD}
P(D) = \frac{1}{\sqrt{2\pi} \delta_D D} e^{-\frac{\ln^2(D/D_0)}{2\delta_D^2}}
\end{equation}
with parameters $D_0$ and $\delta_D$.
The parameter $D_0$ in Eq.~(\ref{PD}) is the median of $P(D)$ and not the average diameter, which reads $D_0 e^{\delta_D^2/2}$.
Moreover, $\delta_D$ is a dimensionless parameter related to the standard deviation (SD) by $\text{SD} = D_0 e^{\delta_D^2/2}\sqrt{e^{\delta_D^2}-1}$. 
Defining the root-mean-square deviation as the ratio between the standard deviation and the mean diameter, we therefore get 
\begin{equation}\label{rms}
\text{r.m.s.} = \sqrt{e^{\delta_D^2} - 1} \simeq \delta_D
\end{equation}
Hence, $\delta_D$ is a measure of the root-mean-square deviation.
Samples with $\delta_D \lesssim 0.1$ (10\%) are usually considered monodisperse in the MNP synthesis literature.

From the distribution of diameters $P(D)$, we may also look at the distribution of volumes $P(V)$. 
For this purpose, the  lognormal distribution turns out to be quite useful since it satisfies  the following very unique  property:
If $D$ is lognormal, then so will $a D^k$, with parameters $a D_0^k$ and $|k| \delta_D$. 
This means that the distribution of volumes $P(V)$ will also be given by a lognormal distribution, which for spherical particles will have parameters $V_0 = \frac{\pi}{6}D_0^3$ and $\delta_V = 3 \delta_D$ (for non-spherical particles only  $V_0$ must to be modified).  That is, the dispersion parameter of the volume distribution is three times larger than the dispersion parameter of the diameter distribution. 
Note that this property is \emph{not} satisfied by other distributions, such as the Gaussian or the Gamma distributions.

Assuming that the anisotropy constant $K$ is the same for all particles, the energy barrier $\theta = KV/k_B$ will, for the same reason, also be given by a lognormal distribution
\begin{equation}\label{Ptheta}
P(\theta) = \frac{1}{\sqrt{2\pi} \theta\delta_\theta} e^{-\frac{\ln^2(\theta/\theta_0)}{2\delta_\theta^2}}
\end{equation}
with parameters 
\begin{equation}\label{theta_parameters}
\theta_0 = \frac{K V_0}{k_B},\qquad \delta_\theta = \delta_V = 3 \delta_D
\end{equation}
The physical meaning of $P(\theta)$ is that, if you randomly draw a particle from your sample, the probability that it will have an energy barrier between $\theta$ and $\theta + \ud \theta$ is $P(\theta)\ud \theta$. 
All our calculations will be done using $P(\theta)$.
The situation where $K$ is not constant will be discussed in Sec.~\ref{ssec:K}.

In order to model the magnetic properties of samples with a size distribution, we must average results  such as Eq.~(\ref{CI}),  over $P(V)$ or $P(\theta)$.
However, this procedure involves a subtle point which has caused considerable confusion in the past. 
It is related to the distinction between the number distribution of volumes, which is the quantity $P(V)$ obtained from TEM, and another distribution called the volume distribution of volumes. 
Although subtle, this point determines the form of the static susceptibility $\chi_0$ in Eq.~(\ref{chi_0}). 
And, if one wishes, to compare AC susceptibility experiments with TEM data, it is essential that the correct form of $\chi_0$ be used. 
For a thorough discussion  clarifying this point, see Ref.~\onlinecite{El-Hilo2012a,*El-Hilo2012}.
We present here a complementary discussion, based on physical arguments.

The main idea is that in an actual experiment, the signal picked up by the magnetometer is always proportional to the magnetic moment of the particles, never the magnetization. 
Thus, even though it is customary to work with magnetization when developing theoretical models, the correct quantity to be averaged when considering a size distribution is always the magnetic moment.

Consider any type of experiment, static or dynamic, and let $\mu(V)$ be the magnetic moment of a particle whose volume is $V$ (by $\mu$ we mean the component of the magnetic moment vector being measured by the pick-up coil). 
Also, suppose that in our sample there are $N_1$ particles with volume $V_1$, $N_2$ particles with volume $V_2$, etc. 
Then the total signal measured by the magnetometer will be 
\begin{equation}\label{mu_tot}
\mu_\text{tot} = \sum\limits_i  \mu(V_i) N_i
\end{equation}
We may also define
\[
p_i = \frac{N_i}{N}
\]
where $N = \sum_i N_i$ is the total number of particles in the sample. 
The $p_i$ represent the probabilities that a particle randomly drawn from the sample will have a volume $V_i$. 
Thus, they correspond exactly to the quantities measured from TEM. 
In terms of them, Eq.~(\ref{mu_tot}) becomes 
\[
\mu_\text{tot} = N \sum\limits_i  \mu(V_i) p_i
\]
For the purpose of clarity, we have assumed that the distribution of volumes is  discrete. 
In order to turn it into a continuous representation we simply replace the sum by an integral; viz., 
\begin{equation}\label{mu_tot2}
\mu_\text{tot} = N \int \mu(V) P(V) \ud V
\end{equation}
In conclusion, in order to average a certain property over  $P(V)$ or $P(\theta)$, as obtained from TEM, it is always necessary to average the magnetic moment, and not the magnetization. 
\footnote{
If one insists on averaging the magnetization it becomes necessary to introduce another distribution, called the volume distribution of volumes, which reads
\[
F_i = \frac{V_i N_i}{\sum_i N_i V_i} = \frac{V_i N_i}{V_\text{tot}}
\]
It represents the fraction of the total magnetic volume which has a volume $V_i$. 
If we define the total magnetization as $M_\text{tot} = \mu_\text{tot}/V_\text{tot}$, then it is possible to show that instead of Eq.~(\ref{mu_tot2}), we get
\[
M_\text{tot} = \int M(V) F(V) \ud V
\]
where $F(V)$ is the continuous version of $F_i$. 
We therefore see that the total magnetization may be written as an ``average'' of the magnetization of each particle. 
However, $F(V)$ is not a probability distribution, so using it to compute averages is incorrect (some modifications are required). 
Moreover, it is not the quantity usually obtained from TEM. 
}

%
%
%
%

\subsection{\label{ssec:poliAC}AC susceptibility for samples with a size distribution}

We are now ready to average Eq.~(\ref{CI}) over the size distribution, using Eq.~(\ref{mu_tot2}).
It is more convenient, however, to average over $P(\theta)$ in Eq.~(\ref{Ptheta}). 
Using the definitions $\theta = K V/k_B$ and $\theta_0 = K V_0/k_B$ to write $V$ in terms of $\theta$ we find that 
\begin{equation}\label{1}
\chi'' = N \frac{(M_s V_0)^2}{3k_B T} \int\limits_0^\infty \frac{\theta^2}{\theta_0^2} \frac{\omega \tau(\theta/T)}{1+[\omega\tau(\theta/T)]^2} P(\theta) \ud \theta
\end{equation}
This result is plotted in Figs.~\ref{fig:example}(b)-(d)  for several values of $\delta_D$ by numerically solving the integral. 
In these figures it is possible to see the gradual effect which an increasing  size dispersion has on the general shape of the curves.

Eq.~(\ref{1}) is exact, but not very convenient to work with. 
An approximate formula may be obtained by noting that the function $\frac{\omega \tau}{1+(\omega\tau)^2}$ will be sharply peaked around
\begin{equation}\label{sigma_star}
\sigma^* = -\ln(2\pi f \tau_0)
\end{equation}
We may therefore approximate\footnote{The factor of $\pi/2$ must be used to ensure that the approximation leaves the area under the curve unaltered. In fact, we have that $\int_0^\infty \frac{\omega \tau}{1+(\omega\tau)^2}\ud \sigma = 2 \tan^{-1} (1/\omega\tau_0)\simeq \frac{\pi}{2}$. This result will only differ significantly from $\pi/2$ when $\omega\tau_0 \gtrsim 1$, which means frequencies in the order of GHz.
}
 $\frac{\omega \tau}{1+(\omega\tau)^2} \simeq \frac{\pi}{2} \delta(\sigma-\sigma^*)$, which leads to
\begin{equation}\label{2}
\chi'' \simeq N \frac{\pi (M_s V_0)^2}{6 k_B} \frac{(\sigma^* T)^2}{\theta_0^2} P(\sigma^* T)
\end{equation}
Note that for frequencies of usual interest in AC susceptibility we have $\sigma^* > 2$, which is the requirement for the validity of Eq.~(\ref{tau1}).\cite{Coffey2013}$^{,64}$

\begin{figure}[!h]
\centering
\includegraphics[width=0.22\textwidth]{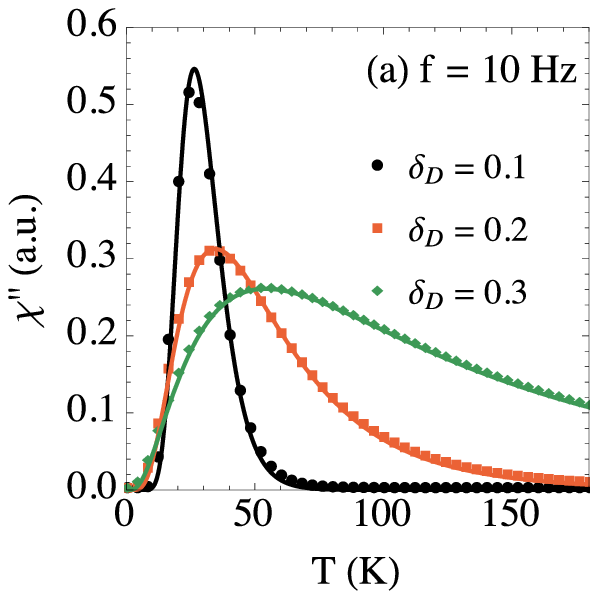}\quad
\includegraphics[width=0.22\textwidth]{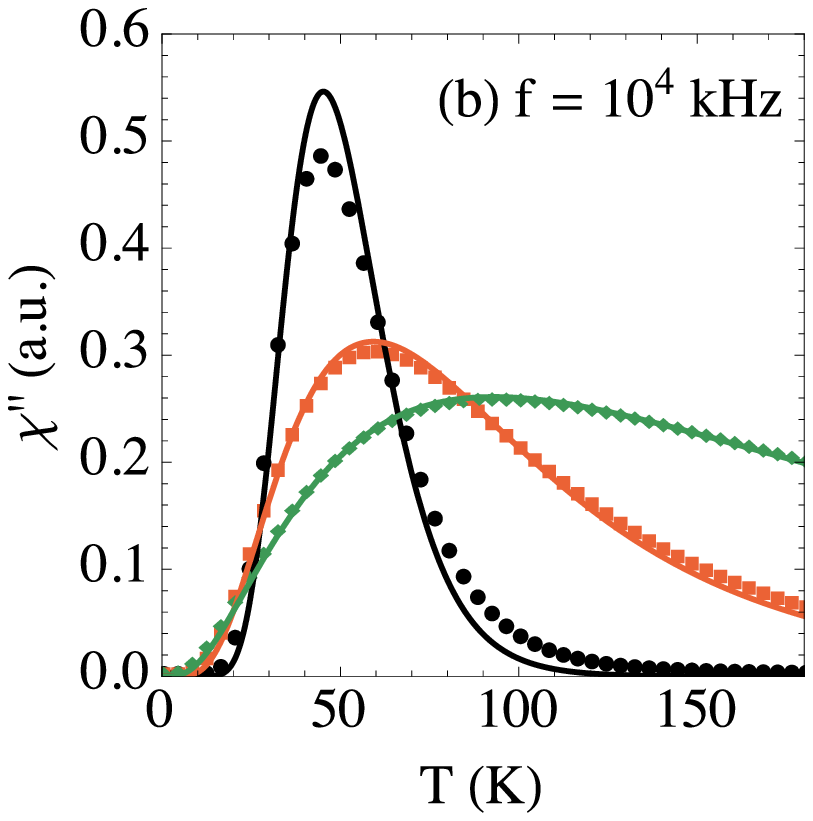}
\caption{\label{fig:compare}(Color Online)
Comparison between the integral formula~(\ref{1}) and the approximate formula~(\ref{2}) for different values of $\delta_D$ and different frequencies.
The other parameters are the same as in Fig.~\ref{fig:example}.
}
\end{figure}

Eq.~(\ref{2}) is valid for any distribution $P(\theta)$.  
Specializing to the case of the lognormal distribution, Eq.~(\ref{Ptheta}),  and defining  
\begin{equation}\label{theta_star}
\theta^* = \sigma^* T = - T \ln(2\pi f \tau_0)
\end{equation}
we finally arrive at\cite{Jonsson1997}
\begin{equation}\label{3}
\chi'' =  c \frac{ \theta^*}{\theta_0^2} \exp{\Bigg\{-\frac{\ln^2(\theta^*/\theta_0)}{2\delta_\theta^2}\Bigg\}}
\end{equation}
where $c = \sqrt{\frac{\pi}{2}} N (M_s V_0)^2/(6k_B \delta_\theta)$ is  a positive  constant.
The reason why $\theta_0^2$ was not included in $c$ is because it is the only term which depends on $K$ and below,  in Sec.~\ref{ssec:max}, we will consider samples which have a distribution of $K$ values. 
In Figs.~\ref{fig:compare} we compare Eq.~(\ref{1}) with Eq.~(\ref{3}). 
It is found that the agreement is overall very good, even for samples with a low dispersion. 
Slight discrepancies are only observed for monodisperse samples ($\delta_D = 0.1$) at high frequencies ($f = 10$ kHz).

Differentiating Eq.~(\ref{3}) with respect to $\theta^*=\sigma^* T$ we find that the maximum of $\chi''$ occurs at $\theta_0 e^{\delta_\theta^2}$. 
This implies the relation
\begin{equation}\label{arrhenius2}
-\ln(2\pi f) = \ln(\tau_0) + \frac{\theta_0 e^{\delta_\theta^2}}{T_\text{max}}
\end{equation}
which shows that even for polidisperse samples an Arrhenius plot will still give a straight line. 
However, the slope of the line is  given by $\theta_0 e^{\delta_\theta^2}$. 
For typical samples we have $\delta_D \sim 0.25$, which leads to   $\delta_\theta = 3\delta_D \sim 0.75$ and thence $e^{\delta_\theta^2}\sim 1.75$. 
This shows that the Arrhenius plot overestimates the anisotropy constant $K$ by a factor which can be almost of the order of 2 in certain cases.
\footnote{
Coincidentally, the volume distribution of volumes is also lognormal,\cite{El-Hilo2012a,*El-Hilo2012} with parameter $\theta_0 e^{\delta_\theta^2}$.
This may lead to some confusion concerning the Arrhenius plot since, if one uses the volume distribution of volumes instead of the number distribution of volumes, it will appear as if Eq.~(\ref{arrhenius}) were valid, even for polidisperse samples. 
However, the quantity $\theta_0$ appearing in the volume distribution of volumes is not $K V_0/k_B$ as obtained from TEM. 
}

We finish this section by noting that, in general, $\chi''$ is a function of both $f$ and $T$. 
However, under the approximations that led to Eqs.~(\ref{2}) or (\ref{3}), it turns out that $\chi''$ will only depend on 
the particular combination  $\theta^* = \sigma^* T = - T \ln(2 \pi f \tau_0)$. 
This means that if we plot each $\chi''$ curve as a function of $\theta^*$ instead of $T$, the data for different frequencies should all collapse into a single curve, provided $\tau_0$ is correctly chosen. 
Hence, this method may be used to determine $\tau_0$ and, since it uses the entire $\chi''(f,T)$ data set,  it turns out to be much more precise than the Arrhenius plot.  
Examples of this are given below for our data in Figs.~\ref{fig:fabi_data} and \ref{fig:irene_data} or, e.g.,  in Refs.~\onlinecite{Masunaga2009,*Masunaga2011b} and \onlinecite{Jonsson1997}.
The collapsed data will  be described by Eq.~(\ref{3}), which depends only on  $\delta_\theta$, $\theta_0$ and $c$.
But $\delta_\theta = 3\delta_D$ is already known from TEM, so that the only two fit parameters are $c$ and $\theta_0$. 
From Eq.~(\ref{theta_parameters}), $\theta_0 = KV_0/k_B$ and $V_0$ is also known from TEM, so that we may extract the anisotropy constant $K$.

%
%
%
%

\subsection{\label{ssec:max}Variation of $\cm$ with frequency}

The usual Arrhenius analysis of AC susceptibility data focuses on the temperature $T_\text{max}$ where  $\chi''$ is a maximum.
We now argue that valuable information concerning the dipolar interaction is contained in the \emph{height} of these peaks. 
Let us denote the maximum of each $\chi''$~vs.~$T$ curve as $\cm$. 
In Fig.~\ref{fig:example} we see two possible behaviors: For monodisperse samples, $\cm$ decreases with $f$, whereas for polidisperse samples, $\cm$ tends to become roughly constant, independent of $f$.
These are the only two possibilities predicted by models of non-interacting particles, including the full Fokker-Planck description of the stochastic Landau-Lifshitz-Gilbert equation.\cite{Brown1963,*Brown1979a,Coffey2004,Poperechny2010,Landi2011c,*Landi2012b,*Landi2012e}

However, in real samples it is also customary to find situations where $\cm$ \emph{increases} with $f$. 
We argue that this is a signature of a strong dipolar interaction. 
It therefore serves as a simple test to estimate the extent of this interaction in a given sample.
If $\cm$ increases with $f$, the dipolar contribution certainly has a significant effect.
This requires no sophisticated analyses, just a simple glimpse at the $\chi''$ curves for different frequencies (this effect was briefly commented  in Ref.~\onlinecite{Hansen2000a}).
Please note  that this rule  is only valid for the imaginary part $\chi''$. 
It does \emph{not} hold for the real part $\chi'$. 

Our claim is  corroborated by extensive experimental evidence in the literature. 
The most clear examples are Refs.~\onlinecite{Masunaga2009,*Masunaga2011b,Jonsson1998,Djurberg1997,Jonsson2000a,Hansen2000a}, 
which study samples with different concentrations. 
In all cases the results are unambiguous: for diluted samples $\cm$ is either constant or diminishes with $f$, whereas for concentrated samples $\chi_\text{max}''$  increases with $f$. 
Additional examples may also be found in Refs.~\onlinecite{Aslibeiki2010,Bittova2012,Goya2003,Jonsson1998a,Kleemann2001,Monson2013,Parker2008,Roca2012,Winkler2008,Zhang1996} and in 
Figs.~\ref{fig:fabi_data} and \ref{fig:irene_data}.

The dipolar interaction creates a tendency for  $\cm$ to increase with $f$. 
But, if the sample is monodisperse, $\cm$ naturally decreases with $f$. 
Hence, in these cases there is a competition between the two effects and care must be taken in analyzing the results. 
This is well illustrated in Figs. 7 and 8 of  Ref.~\onlinecite{Berkov2001}, where the authors simulate the AC susceptibility response of interacting ideally monodisperse particles using the Stochastic Landau-Lifshitz equation. 
They find that, depending on the several parameters of the sample, $\cm$ may either decrease or increase with $f$. 
In Table~\ref{tab:behavior} we summarize the different possible behaviors of $\cm$ with $f$.

%
%

\begin{table}[!h]
\begin{center}
\caption{\label{tab:behavior}How the maximum of the imaginary part, $\cm$, behaves with the frequency $f$.}
\begin{tabular}{c|c|c}
				& 	Monodisperse 	& Polidisperse 	\\ \hline
Non-interacting		&	Decreases	&	Constant	\\
Interacting			&	Inconclusive	&	Increases	\\
\end{tabular}
\end{center}
\end{table}

%
%
%
%

\subsection{\label{ssec:dip_chi}Formulas for the susceptibility including the dipolar interaction}

We will now consider how  Eq.~(\ref{3}) may be modified  to include models of the dipolar interaction. 
We shall focus on three distinct models: the Vogel-Fulcher approximation\cite{Shtrikman1981}, a mean-field variant of this law developed recently in Ref.~\onlinecite{Landi2013,*Landi2014}  and the Dormann-Bessais-Fiorani (DBF) model.\cite{Dormann1999}
This  will provide us with a tool to include the dipolar effect in the susceptibility analysis and  will also serve to further corroborate our claim on the  dependence of $\cm$ with $f$.
It is important to clarify upfront, however, that due to the enormous complexity of the interaction, any model will always involve an enormous number of approximations  and, inevitably, will only be able to capture a fraction of the real phenomenon. 

\begin{figure}[!h]
\centering
\includegraphics[width=0.22\textwidth]{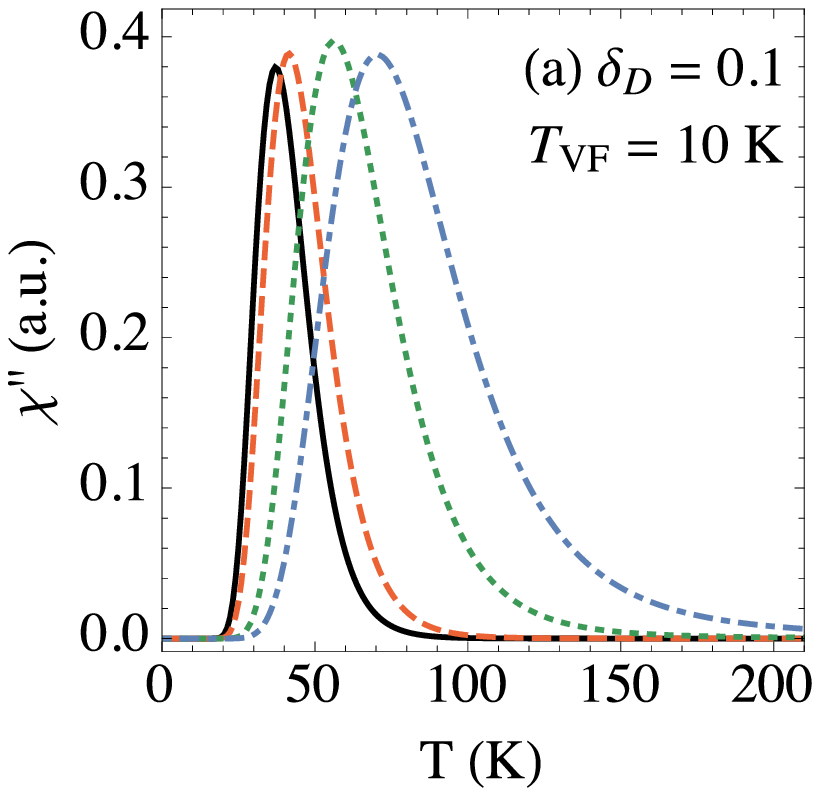}\quad
\includegraphics[width=0.22\textwidth]{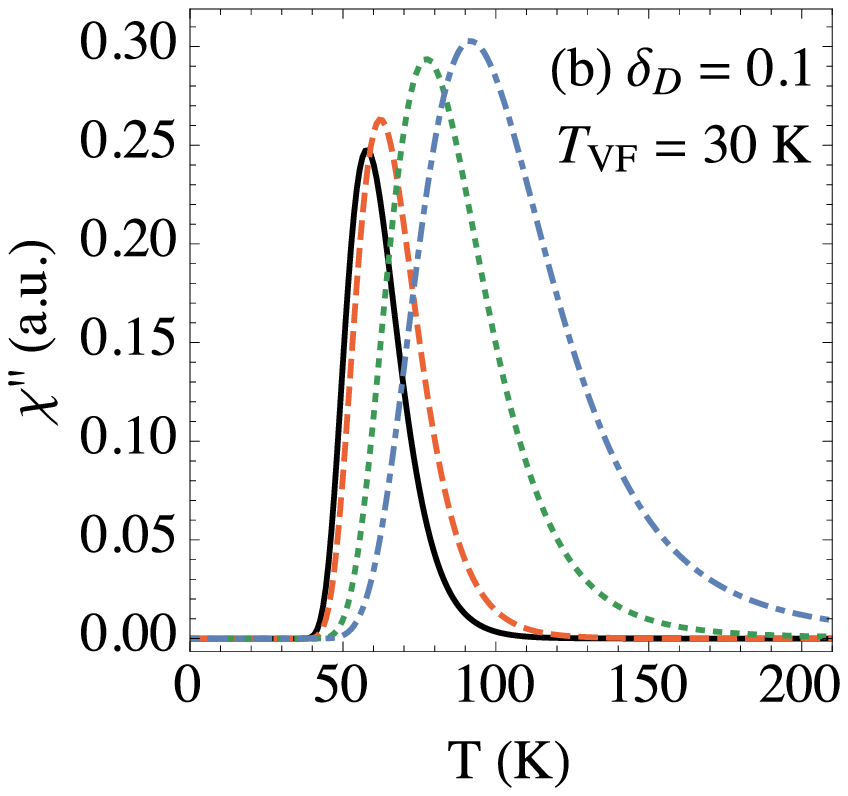}\\
\includegraphics[width=0.22\textwidth]{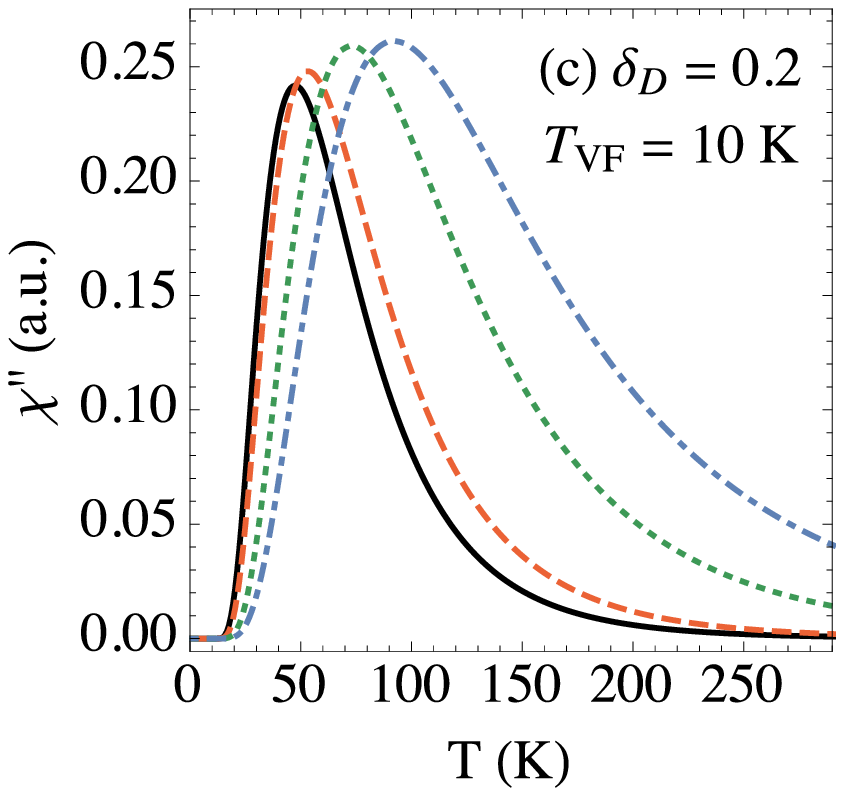}\quad
\includegraphics[width=0.22\textwidth]{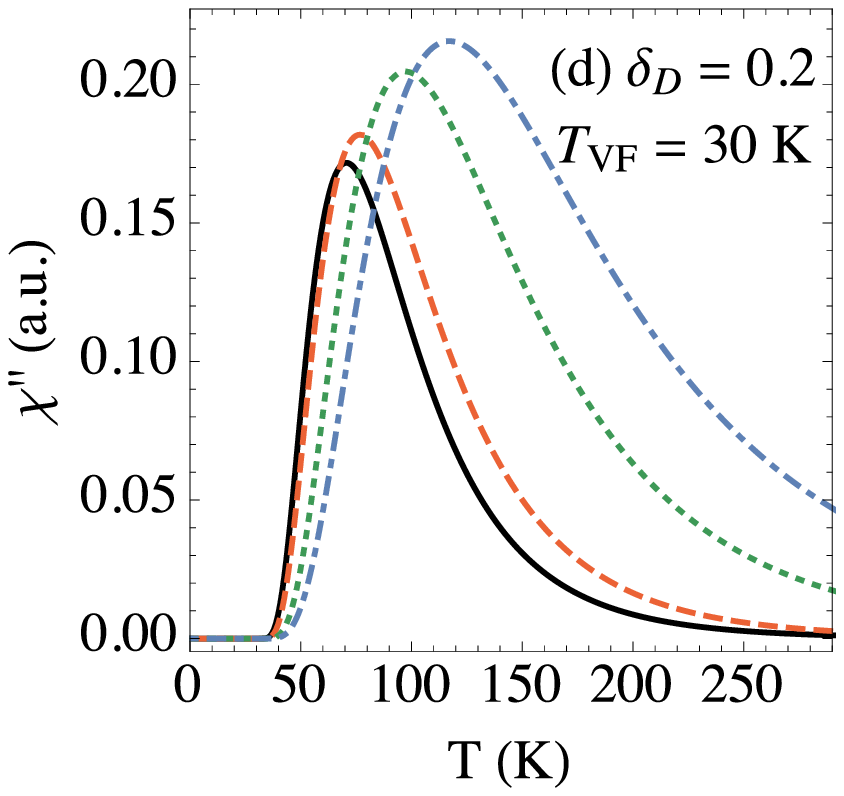}
\caption{\label{fig:VF}
Effect of the Vogel-Fulcher modification, Eq.~(\ref{vf}), in the AC susceptibility curves.
The simulation parameters are shown in the caption of each figure. The other parameters are the same as in Fig.~\ref{fig:example}. 
}
\end{figure}

The most popular model for the dipolar interaction is the Vogel-Fulcher law,\cite{Shtrikman1981} whereby the energy barrier parameter $\sigma$ is modified to account for the dipolar interaction according to 
\begin{equation}\label{vf}
\sigma_\text{VF} = \frac{\theta}{T-T_\text{VF}}
\end{equation}
where $T_\text{VF}$  provides a measure of the strength of the dipolar interaction. 
To study how this modification affects the AC curves we insert  Eq.~(\ref{vf}) in Eq.~(\ref{1}) and compute the integral numerically.
The results are shown in Fig.~\ref{fig:VF} for parameters similar to those in Fig.~(\ref{fig:example}).
It can be seen that the larger  is $T_\text{VF}$, the stronger is the increase of $\cm$  with $f$.
The competition that occurs for monodisperse samples is also evident, particularly when  Fig.~\ref{fig:VF}(a) is compared with Fig.~\ref{fig:example}(b).

In Fig.~\ref{fig:VF}  it can also be seen that  for $T < T_\text{VF}$ the susceptibility is identically zero. 
However, such a zero susceptibility is never observed in real samples, showing an apparent inconsistency of  the Vogel-Fulcher law. 
Namely that the value of $T_\text{VF}$, usually obtained from an Arrhenius plot, is not compatible with the original model from which it was derived. 
This, of course, is just a reflection of the fact that Eq.~(\ref{vf}) is an approximate model to describe an extremely complex effect. 
It therefore can only contain a certain share of the real contribution from the dipolar interaction.

Recently, one of the authors of this paper has worked out a mean-field model to describe the dipolar interaction.\cite{Landi2013,*Landi2014}
The main result of this model is that, instead of Eq.~(\ref{vf}), we should have for sufficiently weak dipolar coupling, 
\begin{equation}\label{mf}
\sigma_\text{MF} = \frac{\theta}{T} + \gamma\left(\frac{\theta}{T}\right)^2
\end{equation}
where\cite{Landi2013,*Landi2014} 

\begin{equation}\label{gamma}
\gamma  = \frac{N}{10} \left(\frac{\mu_0}{4\pi}\right)^2 \left(\frac{\langle \mu^2 \rangle}{K V_0}\right)^2 \left\langle \frac{1}{R^6} \right\rangle \sim T_\text{VF}/\theta_0
\end{equation}
is a dimensionless parameter representing the strength of the dipolar interaction. 
 In this formula $N$ is the number of particles in the sample, $\mu_0$ is the vacuum permeability$, \langle \mu^2\rangle$ is the average magnetic moment squared in the sample and $R$ is  the random variable representing the distance between particles in the sample. 
Eq.~(\ref{mf}) gives results which are qualitatively similar to  those of the Vogel-Fulcher model.
However, it has the advantage that it does not predict a zero susceptibility below a certain temperature.

The mean-field approximation~(\ref{mf}) is also  more convenient than the Vogel-Fulcher model if one wishes to simplify  the integral formula~(\ref{1}) as we did when deriving Eq.~(\ref{3}).
We shall therefore now retrace these steps,  using instead the modified $\sigma_\text{MF}$ of Eq.~(\ref{mf}). 
First, the maximum of $\omega\tau/(1+(\omega\tau)^2)$ will now occur at 
\begin{equation}\label{sigma_starMF}
\sigma_\text{MF}^* = \frac{1}{2\gamma} \Bigg[\sqrt{1- 4 \gamma \ln(2\pi f \tau_0)} - 1\Bigg]
\end{equation}
Moreover, taking into account the relevant experimental ranges for the frequencies $f$,  we may  approximate 
\[
\frac{\omega \tau}{1+(\omega\tau)^2} \simeq \frac{\pi}{2} \frac{1}{\sqrt{1-4 \gamma \ln(2\pi f \tau_0)}} \delta(\sigma-\sigma^*) 
\]
Let us define 
\begin{equation}\label{psi}
\psi(f,\gamma) = \frac{1}{\sqrt{1-4 \gamma \ln(2\pi f \tau_0)}}
\end{equation}
Then Eq.~(\ref{1}) may be approximated to 
\begin{equation}\label{2MF}
\chi'' \simeq N \frac{\pi (M_s V_0)^2}{6 k_B} \psi(f,\gamma) \frac{(\sigma_\text{MF}^* T)^2}{\theta_0^2} P(\sigma_\text{MF}^* T)
\end{equation}
Notice how the dependence of $\cm$ with $f$  now appears explicitly  through the function $\psi(f,\gamma)$. 
The mean-field model therefore predicts that $\cm$ should increase roughly logarithmically (for small $\gamma$) with $f$. A numerical analysis shows a similar dependence  for the Vogel-Fulcher approximation.

We may now finally write the modified version of Eq.~(\ref{3}), which includes the explicit assumption of a lognormal distribution:
\begin{equation}\label{3_MF}
\chi'' =  c\; \psi(f,\gamma) \frac{ \theta^*}{\theta_0^2} \exp{\Bigg\{-\frac{\ln^2(\theta^*/\theta_0)}{2\delta_\theta^2}\Bigg\}}
\end{equation}
where, now 
\begin{equation}\label{theta_star_MF}
\theta^* = \sigma_\text{MF}^* T = \frac{T}{2\gamma} \Bigg[\sqrt{1- 4 \gamma \ln(2\pi f \tau_0)} - 1\Bigg]
\end{equation}
According to these results, we see that in the weakly interacting case [which is where the mean-field approximation~(\ref{mf}) applies] one should be able to perform a data collapse by plotting $\chi''/\psi$~vs.~$\theta^*$.

Both previous models hold for weakly interacting systems. 
In the case of strong interactions, the DBF model\cite{Dormann1999} predicts instead that the relaxation time of the system should be modified according to 
\begin{equation}\label{tau_DBF}
\tau = \tau_0 e^{-n_1}  \exp\bigg\{ \sigma\left( 1+ \frac{n_1 a_1 M_s^2}{K} \right)\bigg\}
\end{equation}
where $n_1$ is the number of nearest neighbors and $a_1$ is the volume concentration of particles in the sample. 
This model therefore predicts two results. 
First,  that $\tau_0$ should be effectively reduced by a factor $e^{-n_1}$. 
This explains the unphysically low $\tau_0$ values which are  sometimes obtained using Arrhenius plots. 
Second, it predicts an increase in the energy barrier $\sigma$ according to
\[
\sigma_\text{DBF} = \sigma\left( 1+ \frac{n_1 a_1 M_s^2}{K} \right)
\]
Notice that the $\sigma$ (or $V$) dependence is linear in this case and not non-linear as in Eqs.~(\ref{vf}) and (\ref{mf}). Consequently, the DBF model would not predict a change in the height of the $\chi''$~vs.~$T$ curves in the strongly interacting regime. 


\subsection{\label{ssec:K}Distribution of anisotropy constants}

As discussed in Sec.~\ref{sec:int}, in addition to the usual contribution from the dipolar interaction, we must also consider the distribution of anisotropy constants caused by the formation of aggregates. 
The particles in a sample are either in free suspension in the fluid or reside in clusters of different sizes.\cite{Castro2008}
And depending on the size of the cluster a particle resides  and in the position of the particle within that cluster, it may experience a different modification to its anisotropy constant $K$.\cite{Jacobs1955,Branquinho2013}
We therefore see that, in addition to the distribution of volumes in a given sample, we should also expect to have a very complex  distribution of $K$ values. 
This effect is likely much more significant than the intrinsic fluctuations due to the crystallinity, shape and surface roughness that exist in every sample. 

To include this effect, we must perform a second average of Eq.~(\ref{3_MF}) over a distribution of $K$.
This distribution is certainly continuous, but we have no information about it to be able to propose an analytical formula. 
Hence, we will for simplicity assume that the distribution of anisotropy constants is discrete. 
That is, we will assume  that in the sample there is a fraction $q_1$ of particles which have an anisotropy constant $K_1$, a fraction $q_2$ with constant $K_2$, etc. 
In order to obtain a tractable equation, we will also assume that $\gamma$ defined in Eq.~(\ref{gamma}) is the same for all $K$, an  approximation  which is justified due to the smallness of $\gamma$. 
We then obtain, instead of Eq.~(\ref{3_MF}),
\begin{equation}\label{4}
\chi'' =  c\; \psi(f,\gamma) \sum\limits_{i=1}^{n_K} q_i \frac{\theta^*}{\theta_{0,i}^2} \exp{\Bigg\{-\frac{\ln^2(\theta^*/\theta_{0,i})}{2\delta_\theta^2}\Bigg\}}
\end{equation}
where $n_K$ is the total number of  distinct $K$ values we wish to consider and 
\begin{equation}\label{theta0i}
\theta_{0,i} = \frac{K_i V_0}{k_B}
\end{equation}
with $V_0$ being known from TEM. 
Eq.~(\ref{4})   can be fitted to the collapsed data, with  parameters $c$, $q_i$ and $\theta_{0,i}$ (the value of $\delta_\theta$ is fixed from TEM).

Each $K_i$ symbolically represents an environment in which the particle may reside  and the $q_i$ represents the fraction of the total population of particles in that environment. 
By ``environment'' we mean, for instance, a cluster of a given size or the position of the particle within a cluster. 
Due to the enormous complexity of the spatial arrangement of the particles in a sample, 
it is  not possible to make a quantitative definition of the type and number of environments. 
Instead, the purpose of this procedure, is to quantify the importance of the dipolar interaction in the sample and the complexity of the particle arrangements.
If a given sample requires a large number of $K_i$ values to be adequately described by Eq.~(\ref{4}), then the dipolar interaction is likely producing a complex modification in the energy landscape of the particles. Moreover, by analyzing the different $K_i$ values obtained for each sample and comparing them with the expected $K_i$ values of a non-interacting ensemble, we may infer the types of modifications brought about by particle aggregation.

%

As a final technical comment, we should mention that in principle $\tau_0$ also depends on $K$, so we should use different $\tau_0$ values for each term in the sum~(\ref{4}). 
However, the influence of $\tau_0$ in the susceptibility is logarithmic and we have found that adding this additional complication produces no improvements whatsoever in the numerical analyzes. 
We have therefore opted to assume a fixed $\tau_0$ for all populations.

%
%
%
%

\section{\label{sec:exp}Experiments}

We now apply Eq.~(\ref{4}) to two distinct samples. 
In Sec.~\ref{ssec:fabi} we study a commercial ferrofluid containing spherical magnetite nanoparticles of roughly 6.4 nm in diameter and in Sec.~\ref{ssec:irene} we study a sample of cubic magnetite nanoparticles with about 13 nm  loaded on the surface of PLGA nanospheres. Both samples are studied under different dilutions.
All non-linear fits to be presented below were performed using a combination of the Nelder-Mead and  Differential Evolution methods. 
Each fit was repeated several times with different initial random seeds and the result which led to the smallest global merit function was used.

\subsection{\label{ssec:fabi}Spherical magnetite nanoparticles}

\begin{figure}
\centering
\includegraphics[width=0.22\textwidth]{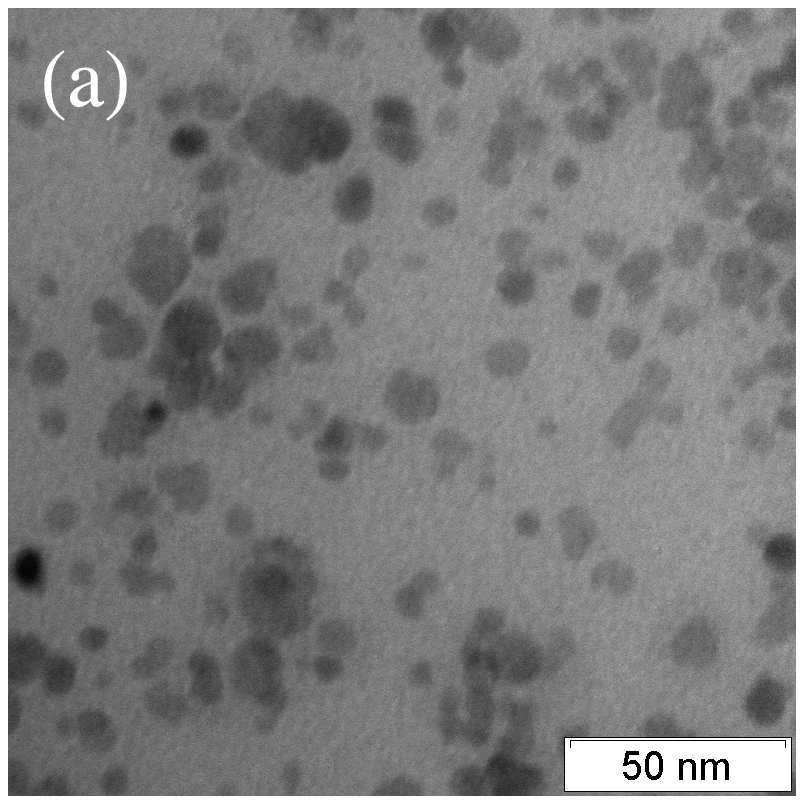}\quad
\includegraphics[width=0.22\textwidth]{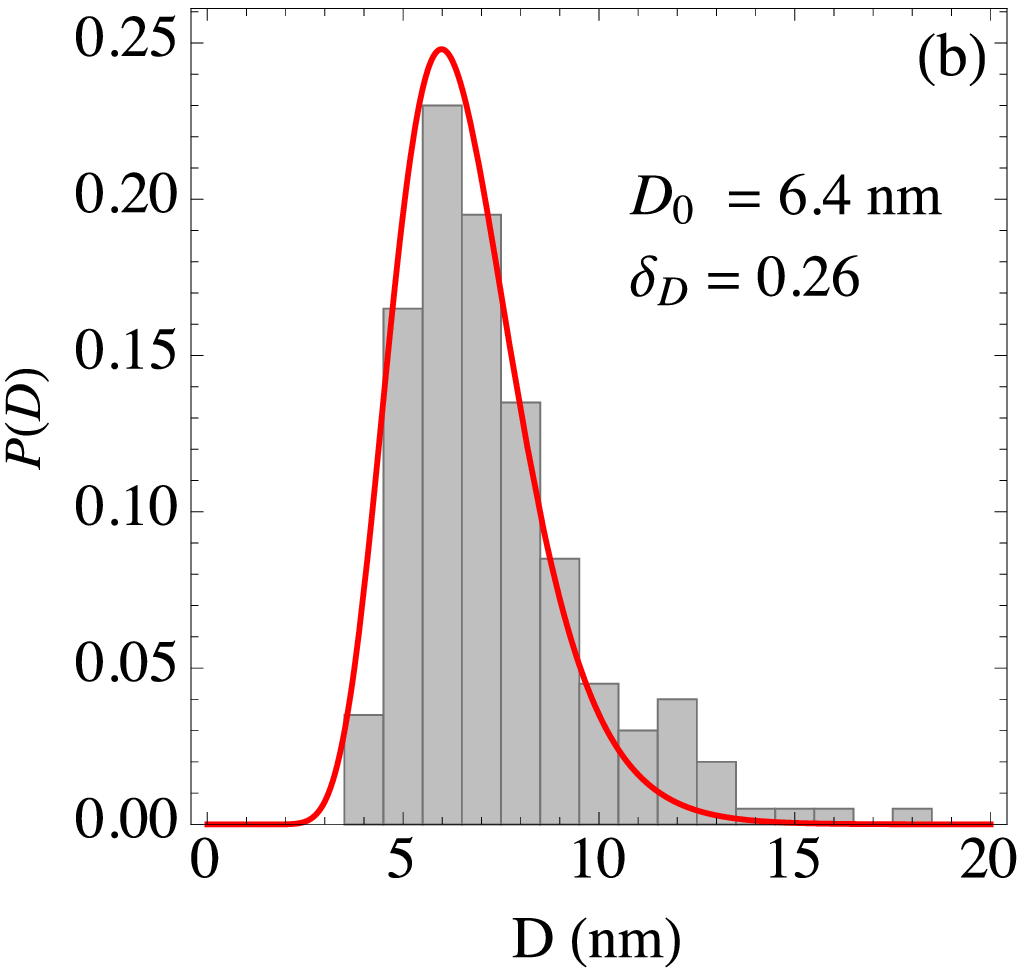}
\caption{\label{fig:fabi_size}
(a) TEM micrograph of the commercial ferrofluid EMG909 from Ferrotec Co. 
(b) Corresponding size distribution and lognormal fit.
Best fit parameters are shown in the figure. 
}
\end{figure}

We begin with the commercial ferrofluid EMG909 from Ferrotec Co, consisting of roughly spherical magnetite particles dispersed in oil. 
A typical TEM micrograph is shown in Fig.~\ref{fig:fabi_size} together with the size distribution histogram and lognormal fit. 
The best fit parameters were $D_0 = 6.4$ nm and $\delta_D = 0.26$. 
Thus, the mean diameter is 6.6 nm and the dispersion of the energy barrier distribution is $\delta_\theta = 3 \delta_D = 0.78$. 

\begin{figure}
\centering
\includegraphics[width=0.22\textwidth]{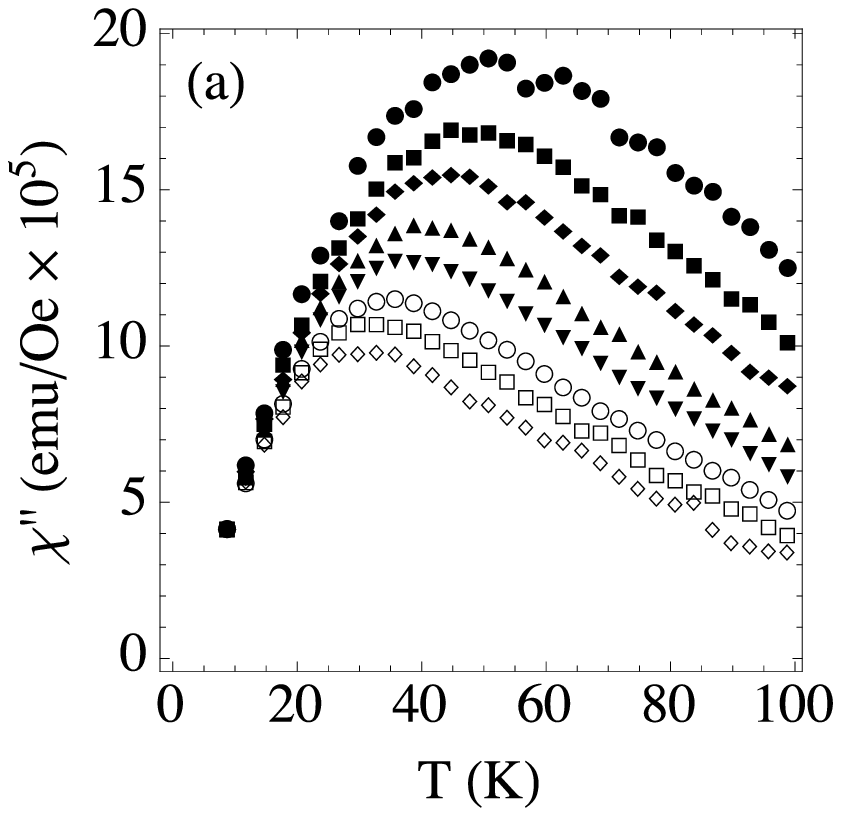}\quad
\includegraphics[width=0.22\textwidth]{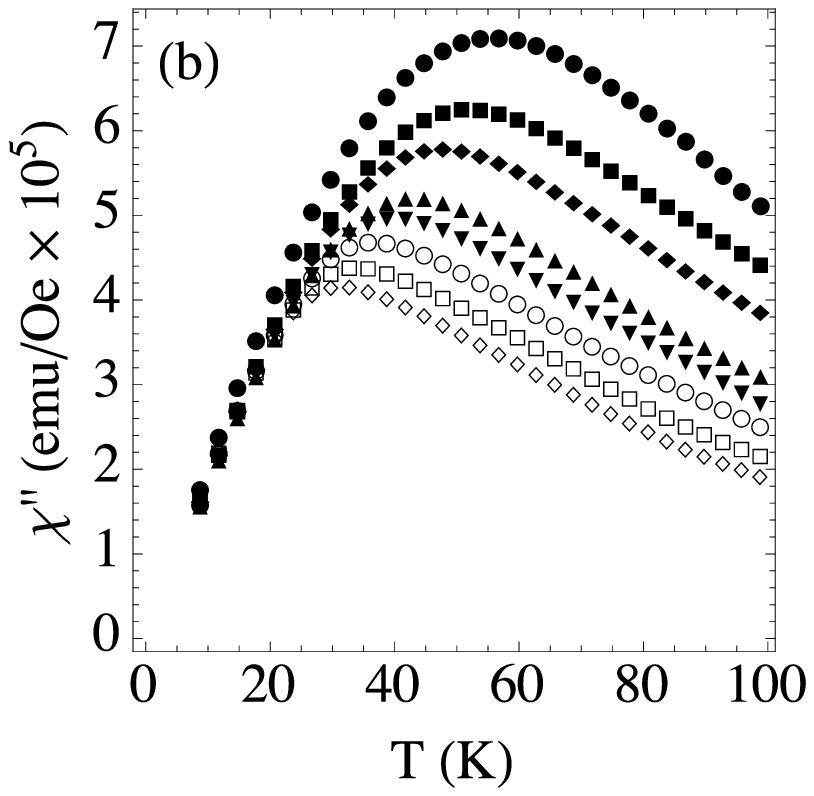}\\
\includegraphics[width=0.22\textwidth]{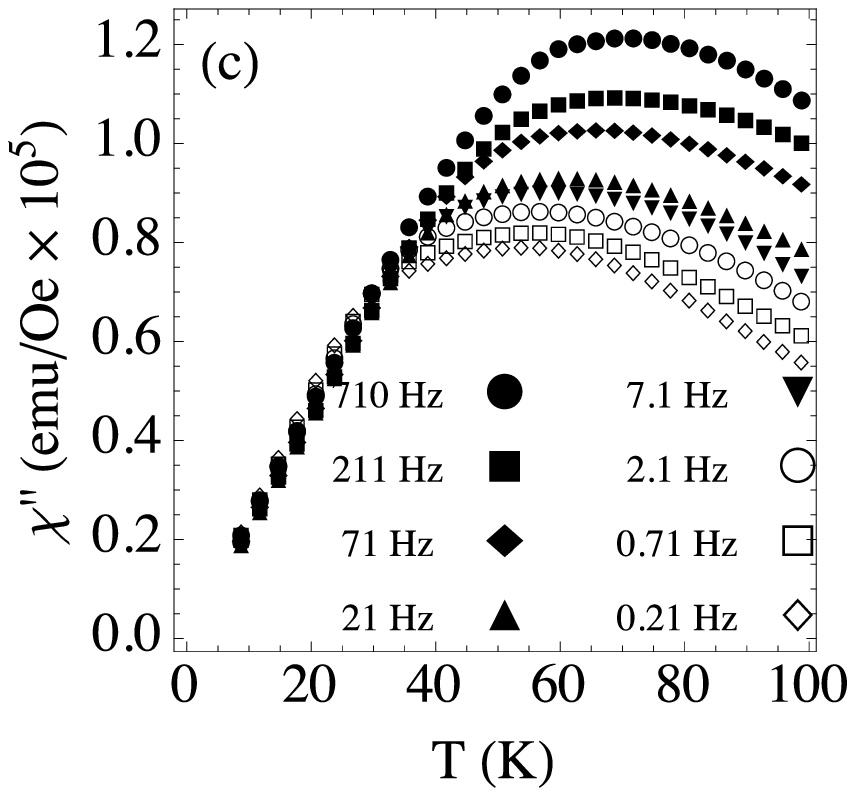}\quad
\includegraphics[width=0.22\textwidth]{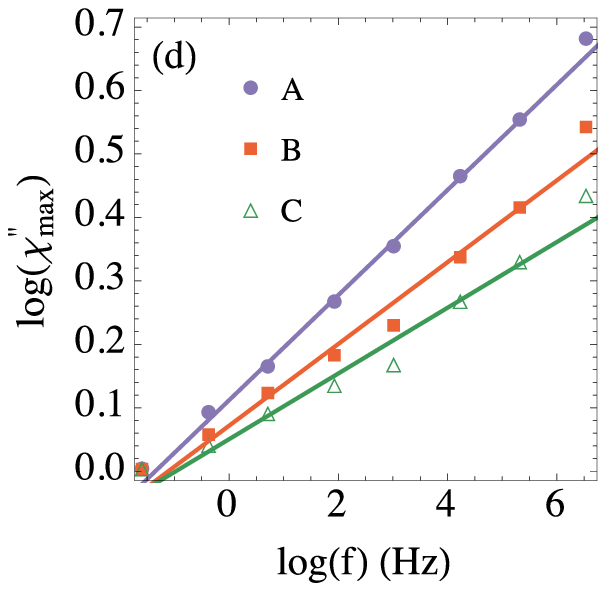}
\caption{\label{fig:fabi_data}
Raw AC susceptibility data for several frequencies for samples (a) A, (b) B and (c) C. 
The volume concentrations of magnetic material were 0.0054\%, 0.54\% and 3.6\% respectively.
Image (d) shows a log-log plot of the maximum of each $\chi''$ curve as a function of the frequency $f$ for the three samples. 
Each curve was normalized for visual purposes in order to start at the same point.
The solid lines correspond to linear fits with exponents 0.083, 0.064 and 0.052 for samples A, B and C respectively. 
}
\end{figure}

We have performed AC susceptibility measurements for three dilutions with volume concentrations of magnetic material of 0.0054\%, 0.54\% and 3.6\%.
We shall henceforth refer to them as samples A, B and C respectively. 
The measurements were performed using a Tesla superconducting quantum interference device (SQUID) from Quantum Design under a field amplitude of 2.0 Oe (0.16 kA/m). 
The raw data for $\chi''$ is shown in Fig.~\ref{fig:fabi_data}(a)-(c).
As can be seen, the maximum height $\chi_\text{max}''$ increases substantially with $f$, indicating the strong presence of the dipolar interaction, even in the most diluted sample. 
This is likely due to the formation of aggregates since the contribution from isolated monomers randomly distributed in the liquid would be very low.
The degree of aggregation may be strongly influenced by several  experimental parameters during the synthesis and stable suspension of the nanoparticles. 
 For instance, an inadequate surface coating may decrease the steric and ionic repulsions, leading to a higher degree of aggregation. 
 Another possibility is the aging of the MNPs, which may lead to a desorption mechanism of the coating molecules.\cite{Bakuzis2013,*Castro2008,*Eloi2010}
The temperature where $\chi''$ is maximum also shifts to higher values as the sample concentration is increased.

In Fig.~\ref{fig:fabi_data}(d) we present a log-log plot of the maximum height, $\cm$,  of each $\chi''$~vs.~$T$ curve as a function of the frequency $f$. 
The linearity of the data in a log-log plot shows that $\cm$ depends on $f$ according a power law behavior 
\begin{equation}\label{scaling_alpha}
\cm \sim f^\alpha
\end{equation}
for some exponent $\alpha$. The values of $\alpha$ obtained from a linear fit were 0.083, 0.064 and 0.052 for samples A, B and C respectively. 
This behavior is quite different from that predicted by the mean-field model [Eq.~(\ref{psi})], whereby $\cm$ should depend only logarithmically on $f$. 
This discrepancy, as already pointed out in Sec.~\ref{ssec:dip_chi}, is actually  expected given that the mean-field model only holds for weakly-interacting particles.

We now attempt to perform a data collapse of the $\chi''$ data. 
To do so, we first normalize all curves to have the same height since this aspect has already been analyzed in Eq.~(\ref{scaling_alpha}). 
Next we must consider the choice of $\theta^*$ to be used. 
For non-interacting samples, we should use Eq.~(\ref{theta_star}) and attempt to obtain a single $\tau_0$ value which adequately fits all three samples.  
Conversely,  for interacting samples we could attempt to use Eq.~(\ref{theta_star_MF}), even though we already know from the height analysis that the latter should not be very precise. 
In this case we should try to obtain a $\tau_0$ value which is the same for all three samples, but  different $\gamma$ values for each sample. 

\begin{figure}
\centering
\includegraphics[width=0.45\textwidth]{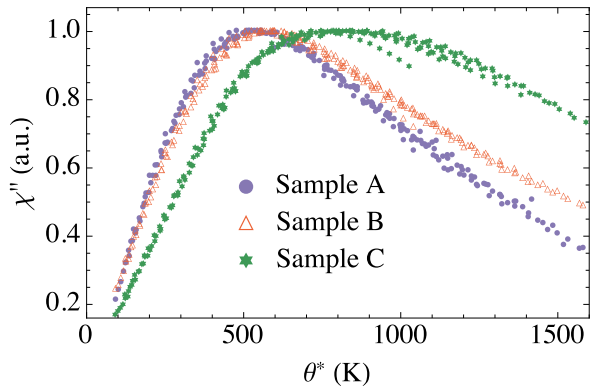}
\caption{\label{fig:fabi_collapse}
Data collapse for samples A, B and C, obtained by plotting $\chi''$ as a function of $\theta^*$, defined either in Eq.~(\ref{theta_star}) or Eq.~(\ref{theta_star_MF}). The results are visually identical in both cases. 
If Eq.~(\ref{theta_star}) is used, we obtain $\tau_0 = 2 \times 10^{-10}$ s, but if Eq.~(\ref{theta_star_MF}) is used we obtain $\tau_0 = 2 \times 10^{-9}$ s and $\gamma = 0.01$, 0.01 and 0.0115 for samples A, B and C respectively. 
}
\end{figure}

The results for both cases are visually identical and for simplicity we only present one,  shown in Fig.~\ref{fig:fabi_collapse}. 
If Eq.~(\ref{theta_star}) is used, we obtain $\tau_0 = 2 \times 10^{-10}$ s, but if Eq.~(\ref{theta_star_MF}) is used we obtain $\tau_0 = 2 \times 10^{-9}$ s and $\gamma = 0.01$, 0.01 and 0.0115 for samples A, B and C respectively. 
We therefore see that by neglecting the dipolar interaction, we underestimate $\tau_0$ by one order of magnitude. 
This, as mentioned above, is a frequent problem in the analysis of interacting particles. 
According to Eq.~(\ref{tau_DBF}) the presence of the dipolar interaction should modify $\tau_0$ by a factor $e^{-n_1}$, where $n_1$ is the average number of nearest neighbors. 
Thus, we may estimate this value by analyzing the ratio between our two estimates of $\tau_0$, the one with the dipolar interaction and the one without it. As a result we get $n_1 \simeq 2.3$.

\begin{figure}[!h]
\centering
\includegraphics[width=0.22\textwidth]{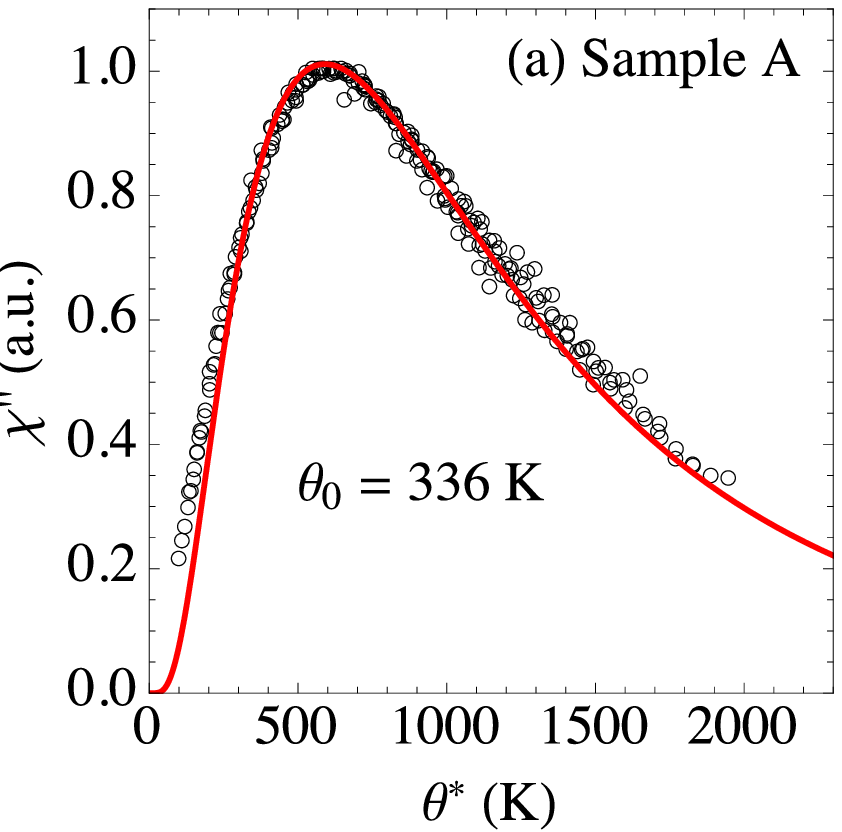}\quad
\includegraphics[width=0.22\textwidth]{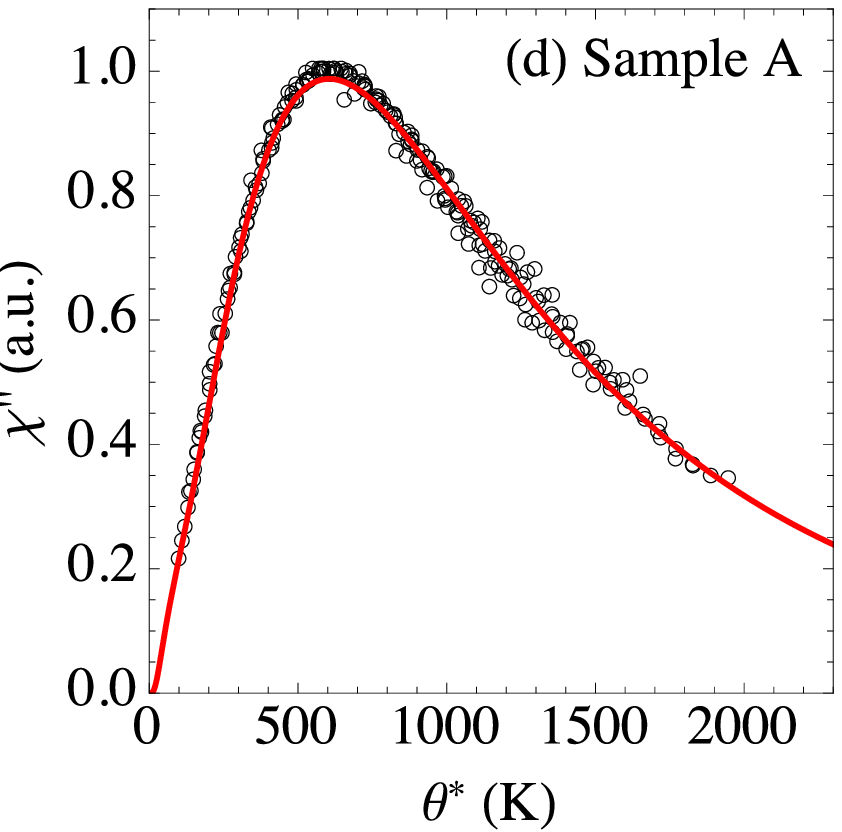}\\
\includegraphics[width=0.22\textwidth]{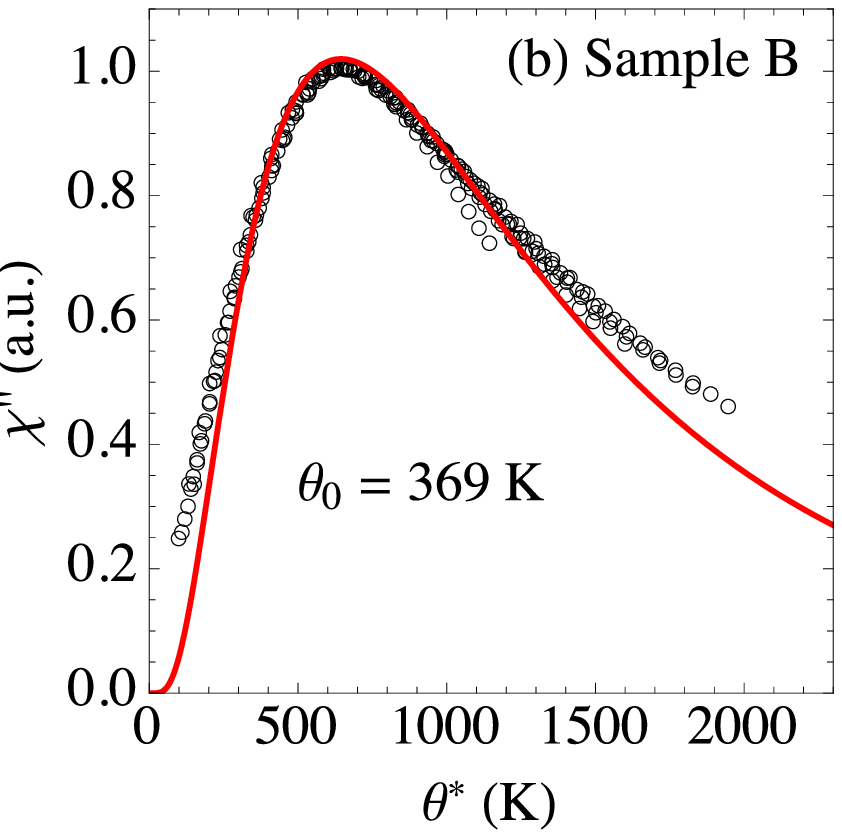}\quad
\includegraphics[width=0.22\textwidth]{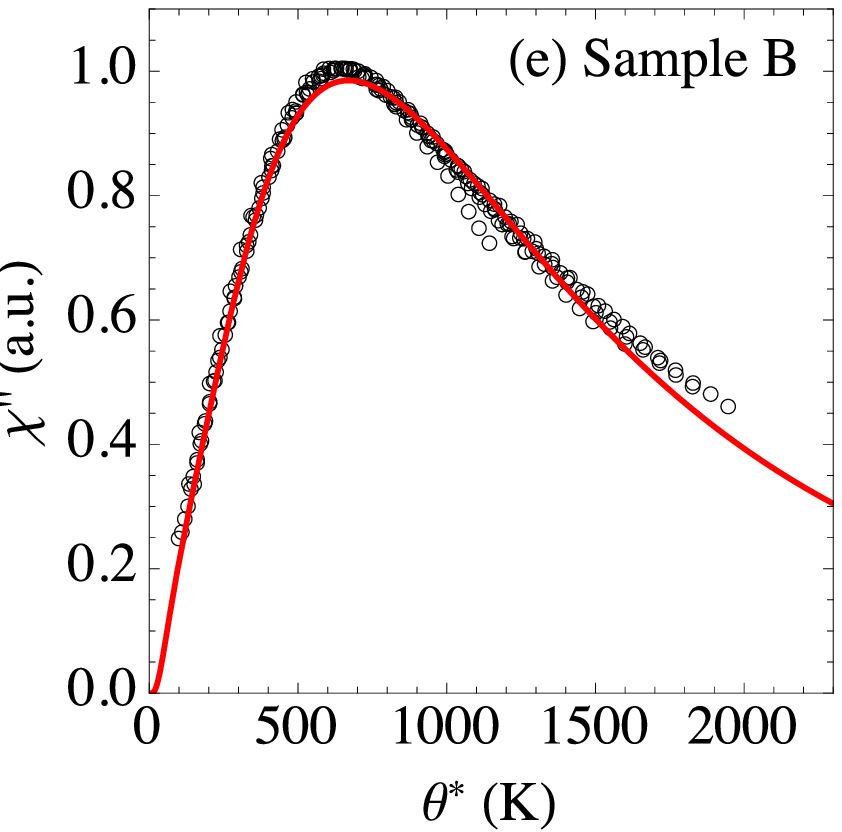}\\
\includegraphics[width=0.22\textwidth]{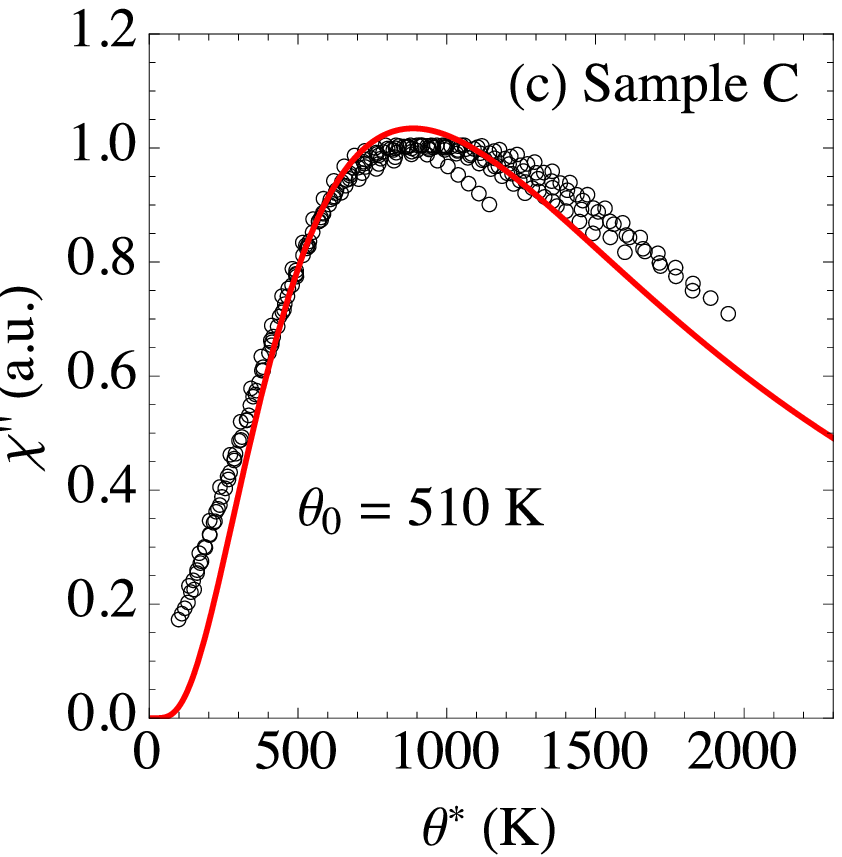}\quad
\includegraphics[width=0.22\textwidth]{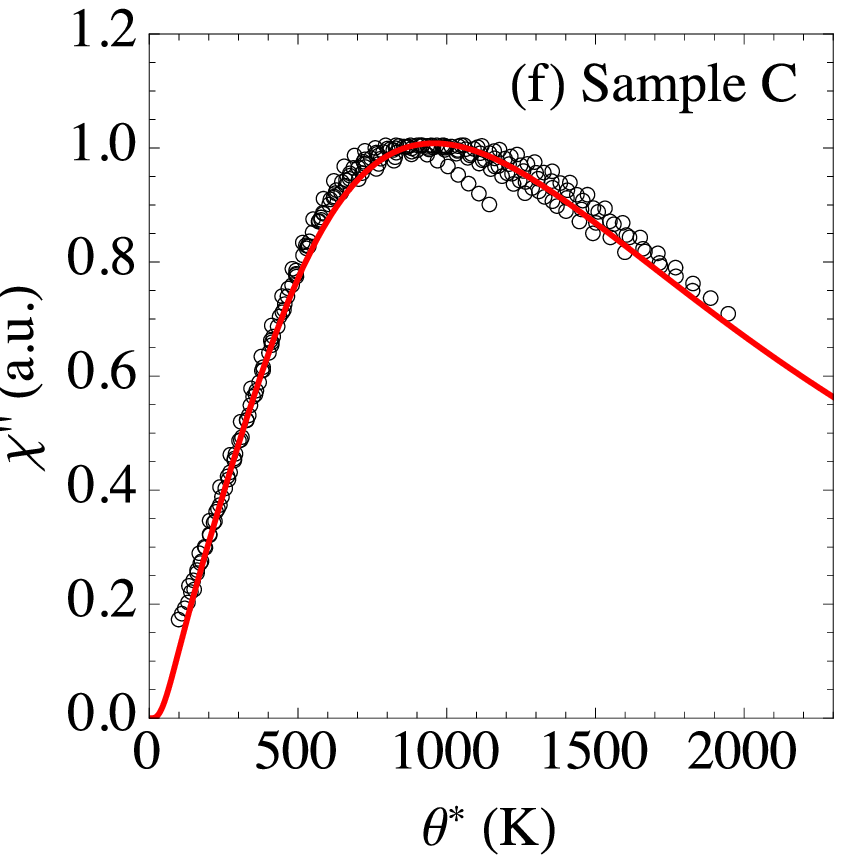}\\
\caption{\label{fig:fabi_fit}
Fit of Eq.~(\ref{4}) to the AC susceptibility data of samples A, B and C. 
Left panel: $n_K = 1$. 
Right panel: $n_K = 2$.
Best fit results for the left panel are shown in each image and for the right panel  in Table~\ref{tab:fabi}.
}
\end{figure}

Once we have obtained a collapsed set of data, the long-range manifestation of the dipolar interaction has been completely accounted for  and we may proceed to use Eq.~(\ref{4}) in order to study the aggregation of particles within the sample. 
Thus, we shall now fit the collapsed data sets using Eq.~(\ref{4}). 
 From TEM we already know that $\delta_\theta = 3 \delta_D = 0.78$ so the only free parameters are $\theta_{0,i}$, $q_i$  and $c$. 
The results for $n_K = 1$ are shown in the left panel of Fig.~\ref{fig:fabi_fit} and the best fit values of $\theta_0$ are presented in each image.
A clear disagreement between the fitted curve and the experimental data can be observed. 
The situation is particularly worse at low temperatures, where the signal from smaller particles is expected to be stronger. 
Note also that this discrepancy is not related to the fact that we fixed $\delta_\theta$; leaving it as a free parameter would not improve the results.

\begin{table}[!h]
\begin{center}
\caption{\label{tab:fabi}Best fit values of Eq.~(\ref{4}) for samples A, B and C when $n_K = 2$. 
}
\begin{tabular}{llcllcll} \hline
 \multicolumn{2}{c}{Sample A} & \qquad \qquad &\multicolumn{2}{c}{Sample B} &  \qquad \qquad & \multicolumn{2}{c}{Sample C}  \\ \hline
 $\theta_{0,i}$ (K) \qquad & $q_i$ &  & $\theta_{0,i}$ (K) \qquad& $q_i$ & & $\theta_{0,i}$ (K) \qquad & $q_i$  \\ \hline
			67		& 	0.026	&	&		89		&	0.039	&	&		137		&	0.04		\\
			387		&	0.974	&	&		438		&	0.961	&	&		625		&	0.96		\\
\end{tabular}
\end{center}
\end{table}

Motivated by this, we now attempt to fit Eq.~(\ref{4}) using  two  $K$ values ($n_K = 2$). 
The results are shown in the right panel of Fig.~\ref{fig:fabi_fit} and the best fit parameters are presented in Table~\ref{tab:fabi}.
As can be seen, in this case the agreement with the experimental data is remarkably good.
From Table~\ref{tab:fabi} we see that both $\theta_{0,1}$ and $\theta_{0,2}$ increase with concentration, indicating that the dipolar interaction shifts the energy barrier distribution of the particles to higher values. 
In Table~\ref{tab:fabi} we also report the fractions $q_i$, showing that the majority of the particles have a high anisotropy. 
Notwithstanding, since Eq.~(\ref{4}) depends on $1/\theta_{0,i}^2$, the small fraction $q_1$ still gives a significant contribution to the AC susceptibility curve, specially at low temperatures. 

\begin{figure}[!h]
 \centering
 \includegraphics[width=0.4\textwidth]{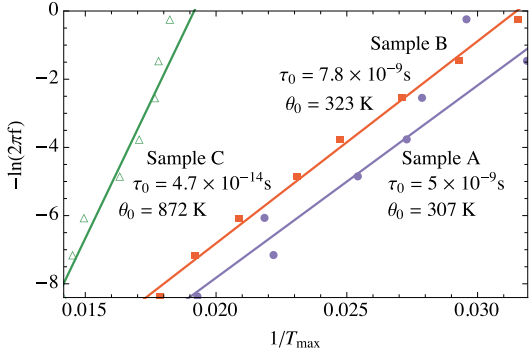}
 \caption{\label{fig:fabi_arr}Arrhenius plots [Eq.~(\ref{arrhenius2})] for samples A, B and C.
 The best fit parameters are shown in the figure. 
In the computation of $\theta_0$ it is necessary to use $\delta_\theta = 3\delta_D = 0.78$. 
}
\end{figure}

 For completeness, we present in Fig.~\ref{fig:fabi_arr} an Arrhenius plot [cf. Eq.~(\ref{arrhenius2})] for the three  samples.
The corresponding best-fit parameters are shown in the figure. 
As can be seen, the Arrhenius plot predicts distinct values of $\tau_0$ for each dilution, which fluctuate substantially. This is related to the strong sensitivity of the Arrhenius plot to experimental uncertainties.  
Moreover, the values of $\theta_0$ are  somewhat similar to those obtained using $n_K = 1$ in the left part of Fig.~\ref{fig:fabi_fit}. 

%
We used a bimodal distribution and obtained the  constants $\theta_{0,1}$ and $\theta_{0,2}$ reported in Table~\ref{tab:fabi}, together with the populations $q_1$ and $q_2$. 
The values of $K_1$ and $K_2$ are then obtained from the relations $\theta_{0,i} = K_i V_0/k_B$. 
For this  sample $k_B/V_0 \simeq 1000$ erg/(K cm$^3$) $ = 0.1 $ kJ/(K m$^3$). 
Hence, using the data from Table~\ref{tab:fabi} we conclude that the  particles in smaller aggregates have anisotropy constants between 6 and 14 kJ/m$^3$.
Conversely, the particles residing in large aggregates have anisotropy constants between 38 and 63 kJ/m$^3$.

%

We therefore observe a roughly 5-fold increase in the effective anisotropy constants between the two populations. 
Such a large difference requires a more careful analysis. 
In Ref.~\onlinecite{Branquinho2013,Bakuzis2013,*Castro2008,*Eloi2010} one of the authors studied a microscopic model of particle aggregation for linear chains, which is the structure expected for colloids within this concentration range. 
It was found that the typical changes in effective anisotropy between chains of different sizes was usually around 2-fold, reaching at most up to 3-fold for certain nanoparticle arrangements. 
The 5-fold change observed in this sample therefore suggests that the size distribution of the particles in  different aggregates may not be the same.
That is,  larger clusters have a tendency to contain larger particles and vice-versa.


\subsection{\label{ssec:irene}Cubic magnetite nanoparticles}

Next we consider the case of PLGA nanospheres loaded with cubic magnetite nanoparticles. 
The confinement of the nanoparticles in the PLGA nanospheres allows maintaining a fixed arrangement of the nanoparticles and prevent uncontrolled aggregation upon increasing concentration. The nanocubes are placed on the surface of the nanospheres, forming quasi 2-dimensional aggregates. 
The details of the synthesis and additional characterization of this sample can be found in Ref.~\onlinecite{Andreu2015}. 

\begin{figure}[!h]
\centering
\includegraphics[width=0.22\textwidth]{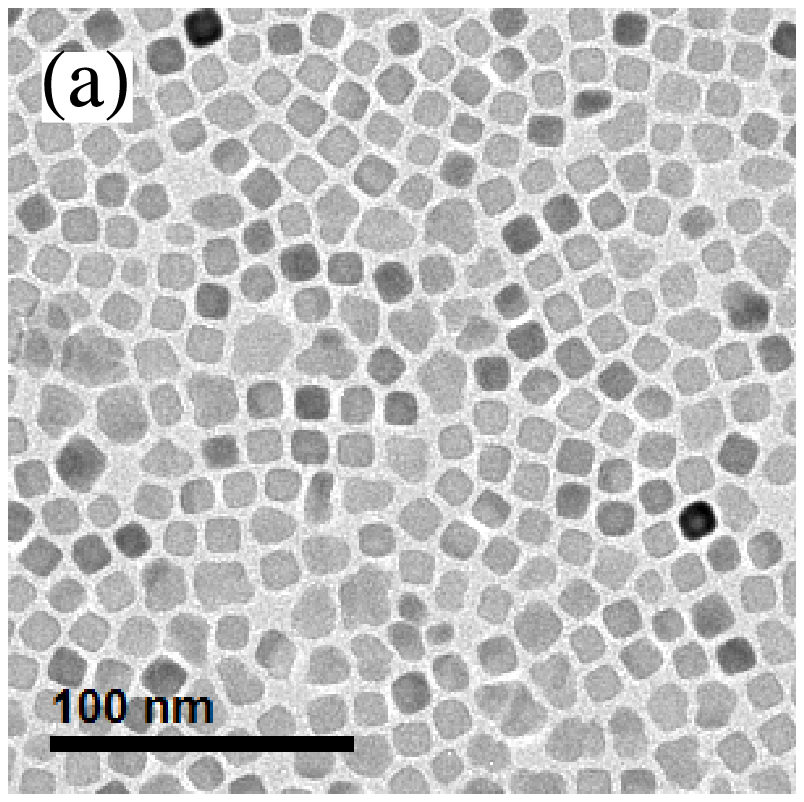}\quad
\includegraphics[width=0.22\textwidth]{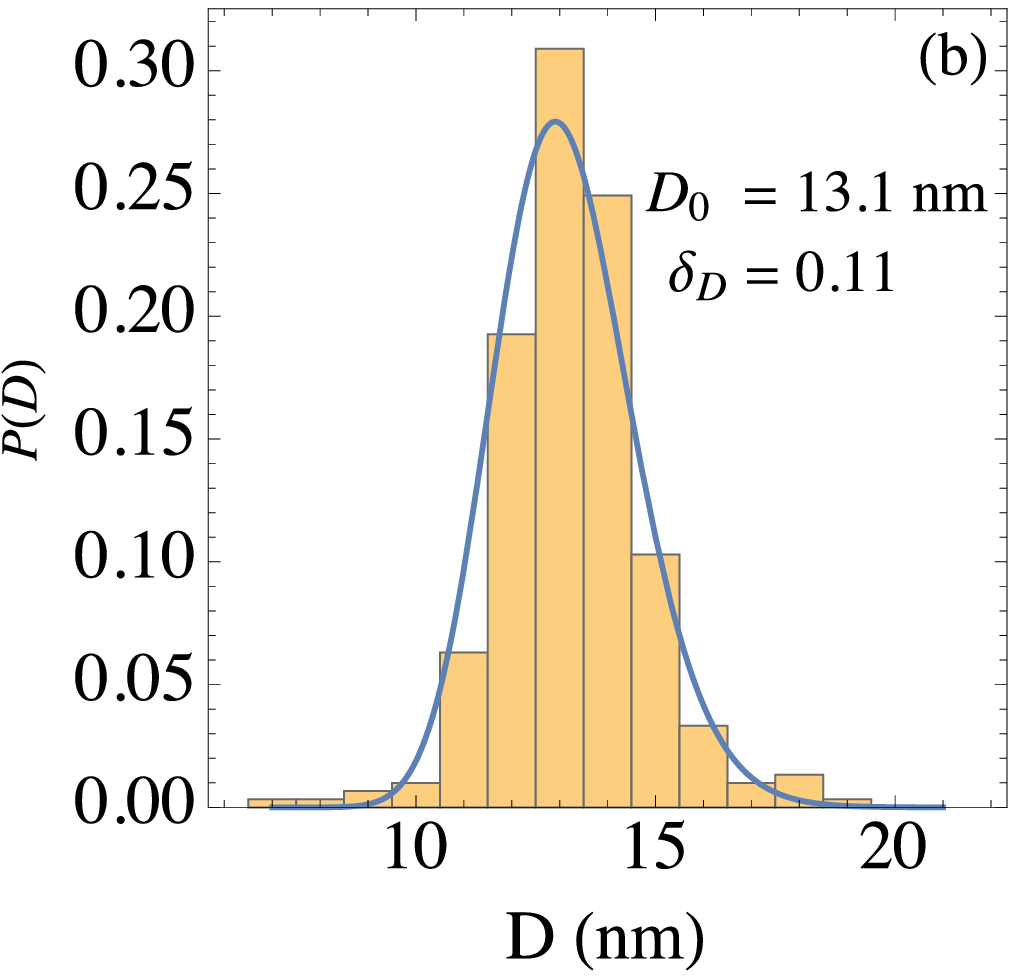}\\
\includegraphics[width=0.47\textwidth]{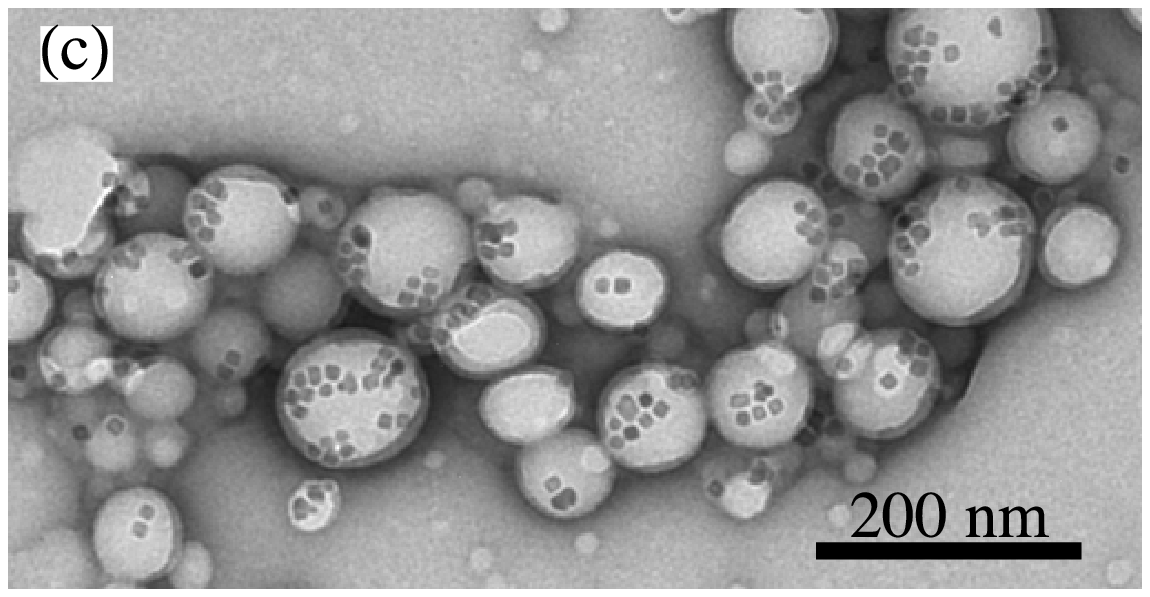}
\caption{\label{fig:irene_size}
(a) TEM micrograph of the cubic magnetite nanoparticles.
(b) Distribution of face diagonals $D$ and corresponding lognormal fit.
The best fit parameters are shown in the figure. 
(c) Example of the nanoparticles loaded into PLGA nanospheres.
}
\end{figure}

All results developed in Sec.~\ref{sec:theory} also hold for cubic particles, provided we now use $V_0 = D_0^3/2^{3/2}$ where $D_0$ is the face diagonal of the cubes. 
Fig.~\ref{fig:irene_size}(a) and (b) show a typical TEM image of the particles together with the size distribution histogram and the lognormal fit.
The best fit parameters were  $D_0 = 13.1$ nm and $\delta_D = 0.11$, showing that the size distribution is very narrow.
Indeed, referring to Eq.~(\ref{rms}), we see that these nano-cubes may be considered as monodisperse.\cite{Andreu2015}
In Fig.~\ref{fig:irene_size}(c) we also show an example of the PLGA nanospheres loaded with MNPs. 

\begin{figure}[!h]
\centering
\includegraphics[width=0.22\textwidth]{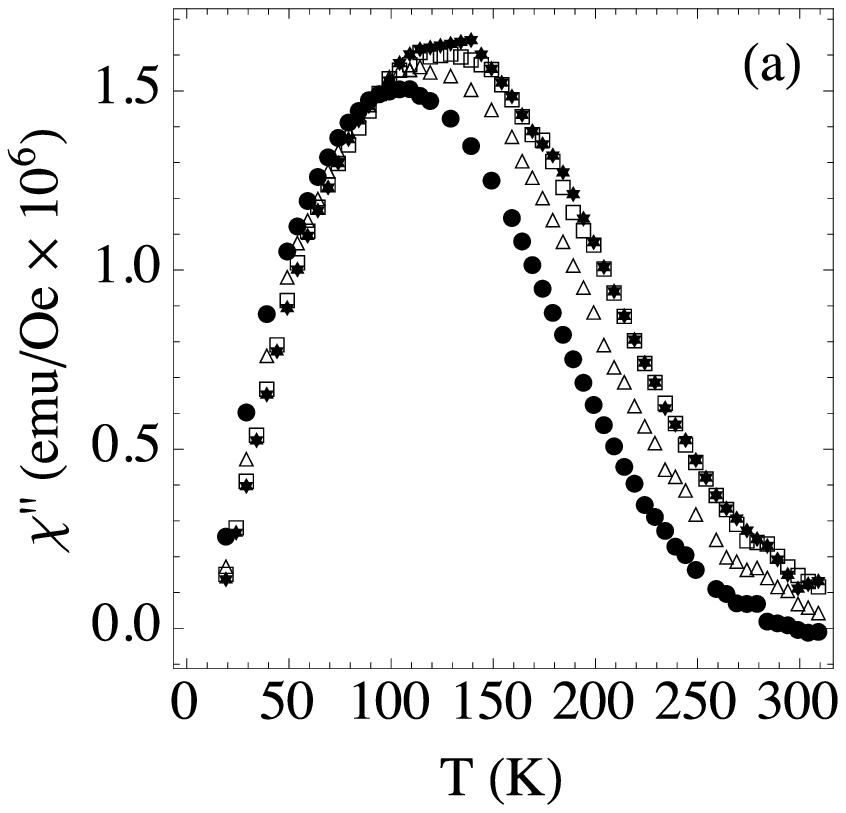}\quad
\includegraphics[width=0.22\textwidth]{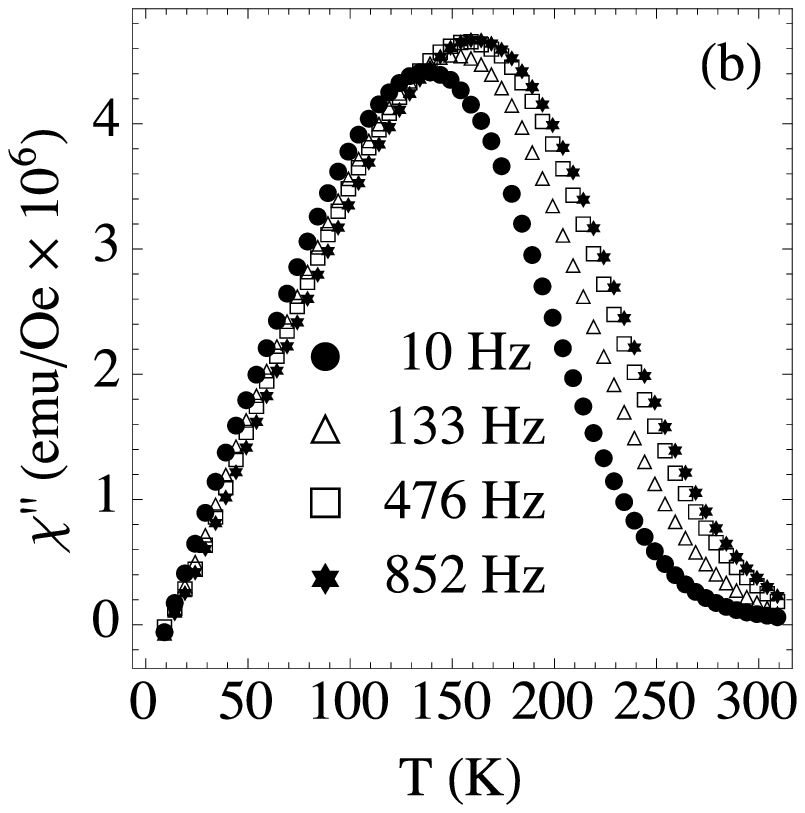}\\
\includegraphics[width=0.22\textwidth]{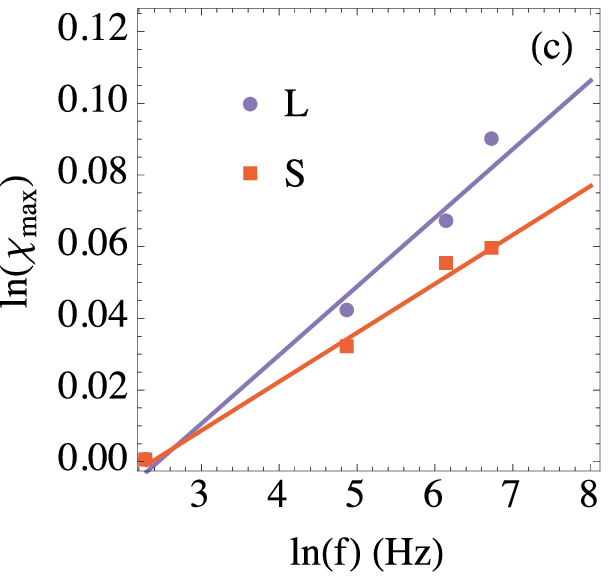}\quad
\includegraphics[width=0.22\textwidth]{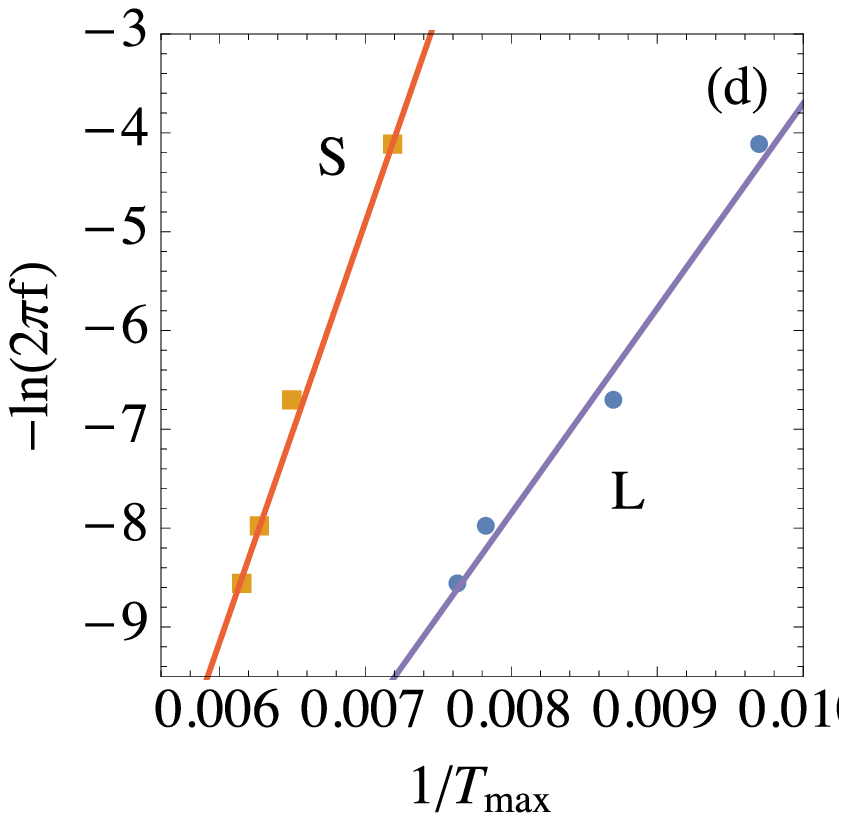}
\caption{\label{fig:irene_data}
(a),(b) Raw AC susceptibility data for samples L (dispersed in water) and S (lyophilized) of cubic magnetic particles. 
The volume concentration of magnetic material was 0.001\% and 0.783\% respectively for samples L and S.
(c) Maximum height $\cm$ as a function of $f$ (normalized by the first frequency point). 
(d) Arrhenius plot with best  fit parameters  $\tau_0 = 2.45 \times 10^{-11} $ s and $\theta_0 = 2073$ K for sample L and $\tau_0 = 8.59 \times 10^{-16}$ s and $\theta_0 = 4256$ K for sample S. 
}
\end{figure}

We have performed AC susceptibility measurements on these nanoparticles in two configurations;
namely with the PLGA spheres dispersed in water and with them lyophilized to form a powder. 
We shall refer to them as samples L and S respectively. 
In these samples, the nanocube arrangements are preserved, but the inter-sphere distance is changed (shorter in S). The volume concentration of magnetic material is 0.001\% and 0.783\% respectively for samples L and S.
In Fig.~\ref{fig:irene_data} we present the raw susceptibility data for both samples and several frequencies, acquired with an MPMS-XL SQUID from Quantum Design under a field amplitude of 2.74 Oe (0.22 kA/m).
As can be seen, in both cases there is a sensitive dependence of $\chi_\text{max}''$ with the height, indicating the presence of a sizable dipolar contribution.
This is due to the compact arrangement of the particles inside the nanospheres.

We begin, as before, by analyzing the dependence of $\cm$ with $f$.
The results are shown in Fig.~\ref{fig:irene_data}(c) where we again observe an approximate power law behavior [cf. Eq.~(\ref{scaling_alpha})], with exponents $\alpha = $ 0.019 and 0.013 for samples L and S respectively. 
We also perform an Arrhenius plot  shown in Fig.~\ref{fig:irene_data}(d). 
The linear fit yields  $\tau_0 = 2.45 \times 10^{-11} $ s and $\theta_0 = 2073$ K for sample L and $\tau_0 = 8.59 \times 10^{-16}$ s and $\theta_0 = 4256$ K for sample S.
We therefore see a clear discrepancy in the $\tau_0$ values, which should be the same since both samples are composed by the exact same particles. 
Moreover, we see that for sample S the value of $\tau_0$ is  clearly unphysical. 
This is again a manifestation of the influence of the dipolar interaction in estimating $\tau_0$. 

\begin{figure}
\centering
\includegraphics[width=0.45\textwidth]{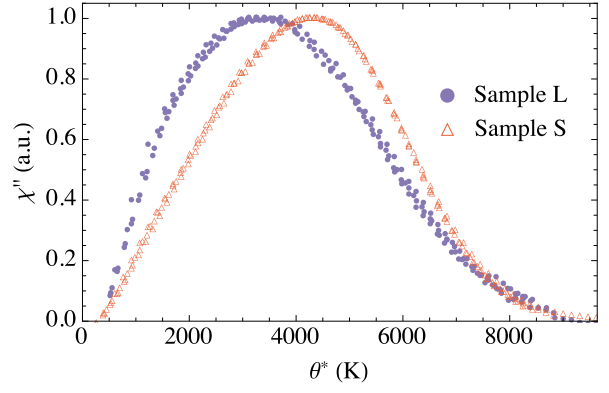}
\caption{\label{fig:irene_collapse}
Data collapse for samples L and S, obtained by plotting $\chi''$ as a function of $\theta^*$, defined either in Eq.~(\ref{theta_star}) or Eq.~(\ref{theta_star_MF}). 
The results are visually identical in both cases. 
If Eq.~(\ref{theta_star}) is used, we obtain $\tau_0 = 5 \times 10^{-16}$ s, 
but if Eq.~(\ref{theta_star_MF}) is used we obtain $\tau_0 = 4 \times 10^{-10}$ s and $\gamma = 0.6$ and 0.9  for samples L and S respectively. 
}
\end{figure}

This problem with $\tau_0$ can be resolved by performing the data collapse of the data using  Eqs.~(\ref{theta_star}) and (\ref{theta_star_MF}). 
The results are again visually identical and are shown in Fig.~\ref{fig:irene_collapse}.
If Eq.~(\ref{theta_star}) is used, we obtain $\tau_0 = 5 \times 10^{-16}$ s, 
but if Eq.~(\ref{theta_star_MF}) is used we obtain $\tau_0 = 4 \times 10^{-10}$ s and $\gamma = 0.6$ and 0.9  for samples L and S respectively. 
We see that neglecting the dipolar interaction yields an unphysical value of $\tau_0$, similar to that of the Arrhenius plot.
But including the mean-field approximation corrects this anomalous behavior. 
The number of nearest neighbors may again be estimated by comparing the  $\tau_0$ values obtained by the two different models and using Eq.~(\ref{tau_DBF}). 
We find  $n_1 \simeq 13.8$, which correlates well with what is expected from the encapsulation of the MNPs in nano-spheres [cf. Fig.~\ref{fig:irene_size}(c)].

\begin{figure}[!h]
\centering
\includegraphics[width=0.22\textwidth]{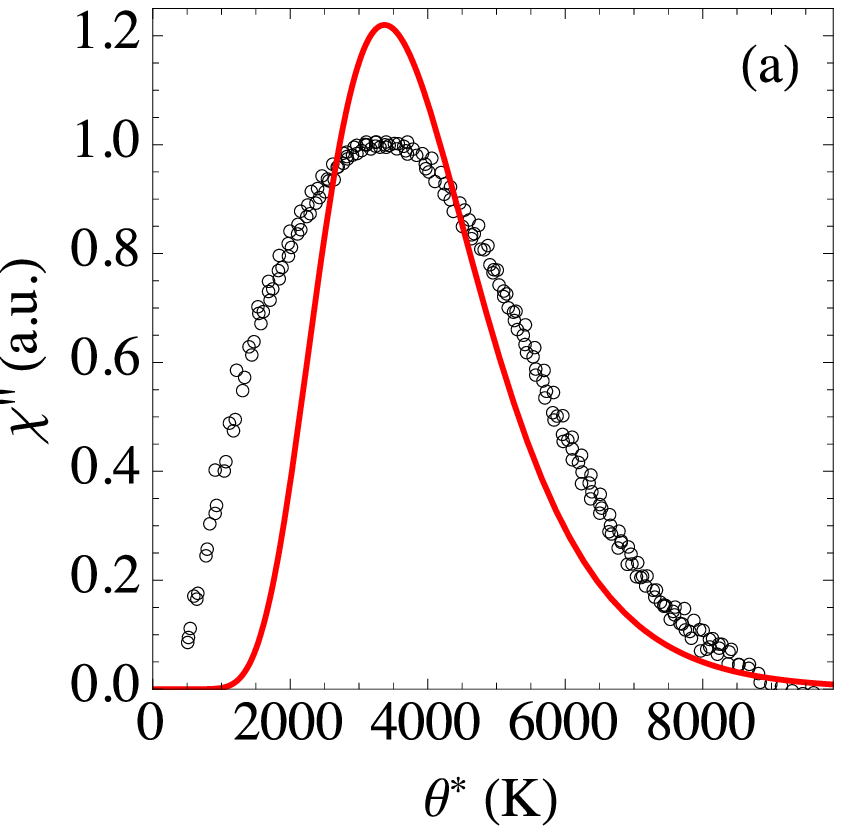}\quad
\includegraphics[width=0.22\textwidth]{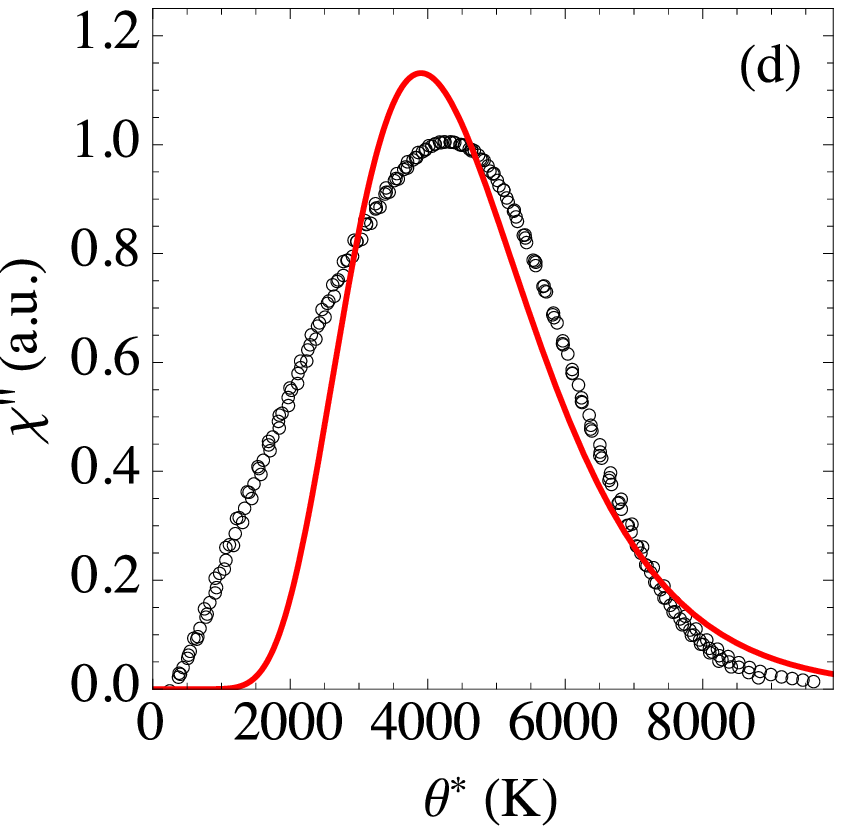}\\
\includegraphics[width=0.22\textwidth]{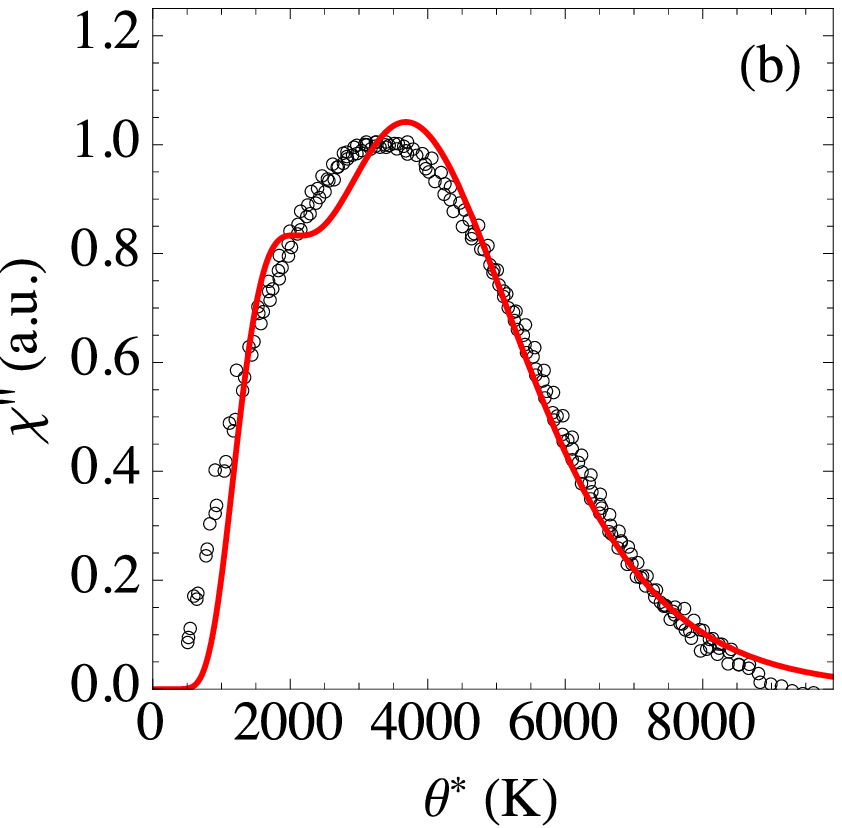}\quad
\includegraphics[width=0.22\textwidth]{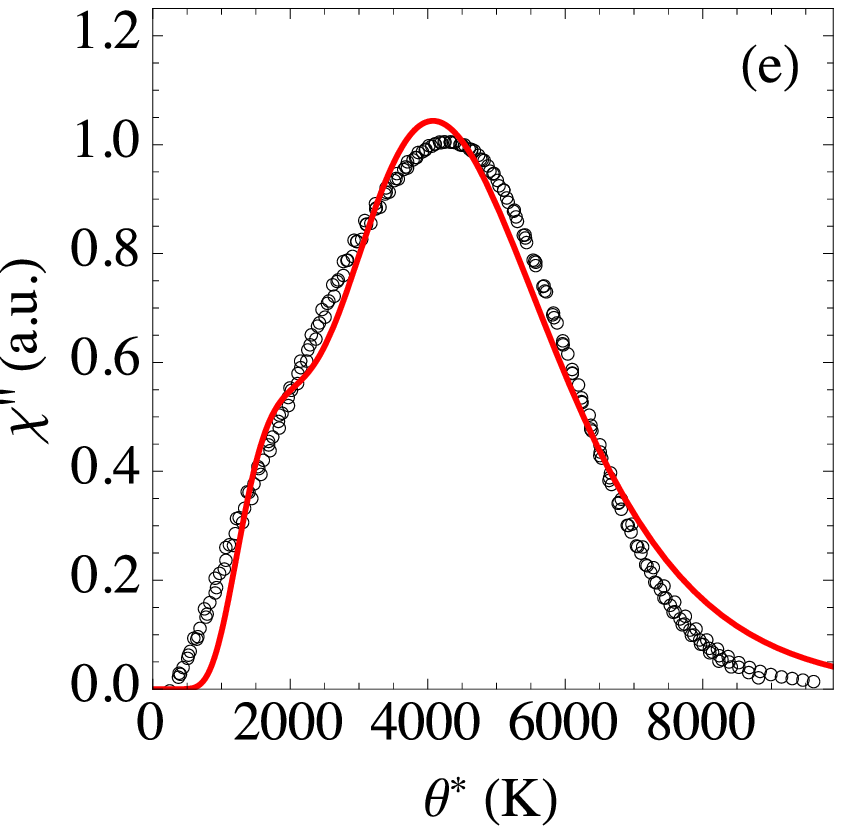}\\
\includegraphics[width=0.22\textwidth]{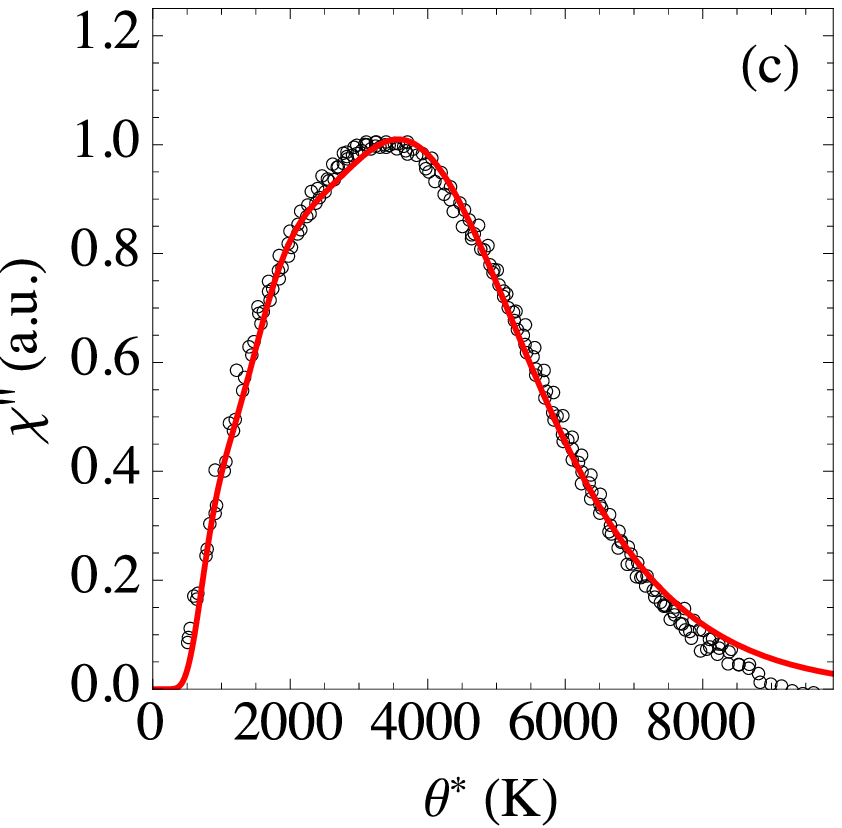}\quad
\includegraphics[width=0.22\textwidth]{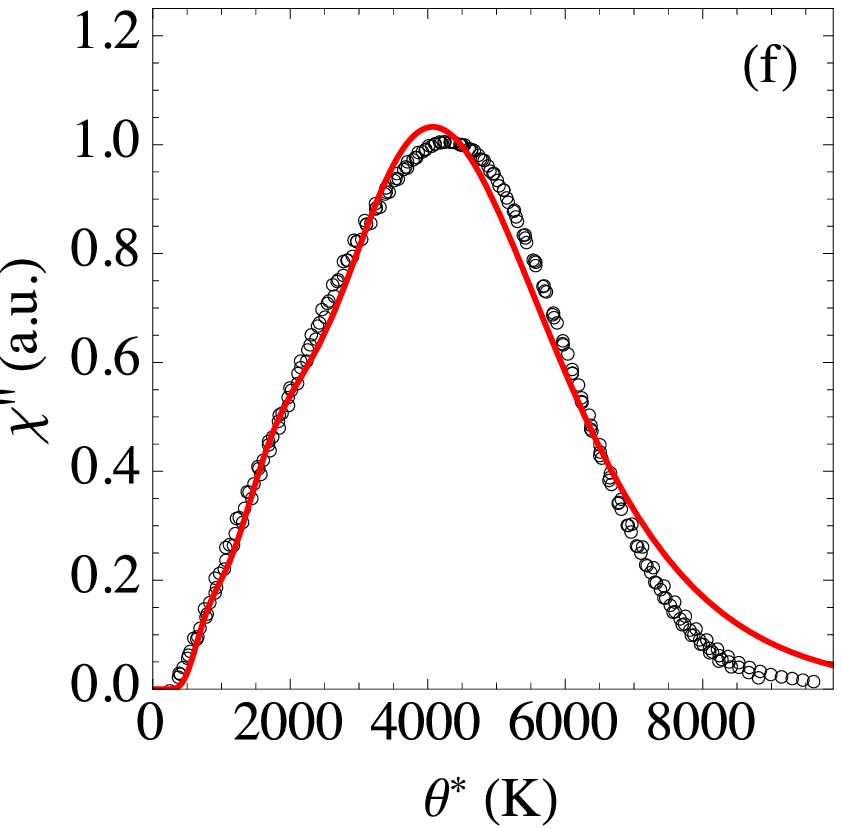}\\
\caption{\label{fig:irene_fit}
Fit of Eq.~(\ref{4}) to the susceptibility data of samples L (left panel) and S (right panel).  
(a) and (d): $n_K = 1$;
(b) and (e): $n_K = 2$;
(c) and (f): $n_K = 3$.
The best fit parameters are summarized in Table~\ref{tab:irene}.
}
\end{figure}

%
%
%
%
We now fit Eq.~(\ref{4}) to the collapsed data in Fig.~\ref{fig:irene_data}(c) using $\delta_\theta = 3 \delta_D = 0.33$, as obtained from TEM. 
The fits were performed using $n_K = 1$, 2 and 3 and the results are shown in Fig.~\ref{fig:irene_fit}, for sample L in the left panel and sample S in the right panel. 
The corresponding best fit parameters are summarized in Table ~\ref{tab:irene}.
As can be seen, there is a strong disagreement between the best fitted function and the experimental data when $n_K = 1$ [images (a) and (d)]. 
The same is true for $n_K = 2$, despite a visible improvement. 
It is only for $n_K = 3$ that the fitted curve begins to resemble the real data. 
As before, this picture does not change if we allow $\delta_\theta$ to  be a free parameter. 
For $n_K = 3$ the results are better for sample L than sample S, in agreement with our intuition that for sample S the distribution of $K$ values should be much more complex due to the increase of inter-sphere interactions. 
Increasing $n_K$ above 3 does not improve the results in any way. 

The results for $n_K = 3$ shown in Table~\ref{tab:irene} are quite interesting to analyze. 
In going from sample L to sample S we see that the energy barriers remain roughly unaltered and only the population of each cluster change.
This is in agreement with the fact that both samples have the same nanoparticle arrangement.
In summary, from TEM we have found that $D_0 = 13.1$ nm (which refers to the face diagonal) and $\delta_D = 0.11$. 
From the data collapse we found $\tau_0 = 5 \times 10^{-16}$ s. 
And, from the fits of the collapsed data, we found that at least three distinct $K$ values are required to correctly describe the data. 
These values of $K$ may be computed using $\theta_{0,i}$ from Table~\ref{tab:irene} 
are roughly 15, 32 and 64 kJ/m$^3$.

\begin{table}[!h]
\begin{center}
\caption{\label{tab:irene}Best fit values of Eq.~(\ref{4}) for samples L and S, with different values of $n_K$. 
Other important parameters are $D_0 = 13.1$ nm, $\delta_D = 0.11$ and $\tau_0 = 5\times 10^{-16}$ s.
}
\begin{tabular}{cllcll} \hline
\qquad\qquad & \multicolumn{2}{c}{Sample L} & \qquad \qquad &\multicolumn{2}{c}{Sample S} \\ \hline
$n_K$ & $\theta_{0,i}$ (K) \qquad & $q_i$ && $\theta_{0,i}$ (K) \qquad& $q_i$ \\ \hline
1 & 3000  & 1 & & 3471 & 1 \\ \hline
2 & 1539 & 0.25 && 1606 & 0.17 \\
& 3444 & 0.75 && 3707 & 0.83 \\ \hline
3 & 916 & 0.064 && 818 & 0.0258 \\
& 1833 & 0.261 && 1751 & 0.166  \\
& 3590 & 0.675 && 3747 & 0.8082 \\ 
\end{tabular}
\end{center}
\end{table}

%
%
%
%

\section{\label{sec:conc}Discussion and conclusions}

The purpose of this paper was to show how AC susceptibility, a technique which is nowadays easily accessible in the laboratory, may be used to extract  information concerning the importance of the dipolar interaction in the sample. 
As we have argued, in samples where the particles are left in fluid suspension or packed inside nano-carriers, the dipolar interaction manifests itself in two separate ways. 
The first is the direct dipolar effect, which can be modeled, for instance, using the Vogel-Fulcher, mean-field or DBF models, as discussed in Sec.~\ref{ssec:dip_chi}. 
In addition, the dipolar interaction also manifests itself by inducing  the formation of clusters of particles within the sample. 
Once in a cluster, the effective anisotropy of a particle will be strongly modified due to its proximity with the other particles.
In this paper we have introduced the idea that this effect can be modeled by including, in addition to the volume distribution usually obtained from TEM, a distribution of anisotropy constants. 
Even though it is not possible to say much about this distribution  in general grounds, we have shown that  by proposing a simple discrete probability distribution this approach can be used to probe the complexity of the dipolar interaction in a sample.

In principle, both contributions should be intertwined. 
However, as we have shown, it is possible to roughly separate the two aspects of the problem by collapsing the data of $\chi''$ obtained for different frequencies. 
The procedure that leads to a collapsed data set involves only the first aspect of the dipolar interaction, whereas the information extracted from the collapsed data involve only the second aspect. 

Practically all models and papers discussing the dipolar interaction focus only on the first aspect. 
It was the purpose of this paper to emphasize the importance of the second aspect as well, specially for samples which are of importance for biological applications.

Both parts of the problem are delicate and must be considered in detail.
For the first, as we have shown, none of the three models considered in Sec.~\ref{ssec:dip_chi} can alone account for all aspects of the experimental data. 
Notwithstanding, a judicious comparison of the results predicted from each model does allow us to extract physically meaningful information.

For the second part, it is important that we  rule out any other possibilities that may explain the discrepancies found when trying to fit just a single $K$ value. 
First, to explain this using only the size distribution would require a large concentration of particles with very small diameter, but no evidence of this was found  in either sample.
Secondly, all the approximations that led us to Eq.~(\ref{3}) have been verified, as already demonstrated in Fig.~\ref{fig:compare}. 
It is also known that Eq.~(\ref{1}) neglects other relaxation times of the particle, such as the transverse relaxation time. 
But these only manifest themselves at very high frequencies, which is not the case here.\cite{Svedlindh1997,Figueroa2013} 
Note also that Brownian relaxation is not allowed since the measurements were performed with the solvents frozen.
Finally, this discrepancy also cannot be explained by a Vogel-Fulcher or mean-field theories discussed in Sec.~\ref{ssec:max}.

It was also the purpose of this paper to call attention to the dependence of $\cm$ with $f$. 
This, as we have shown, provides one with a very simple visual tool to estimate the relevance of the dipolar interaction when analyzing a given sample. 
A summary of the expected behaviors is shown in Table~\ref{tab:behavior}.

In conclusion, we hope to have shown the enormous power of AC susceptibility measurements in extracting information about the dipolar interaction in a sample.
First, by a visual analysis  $\cm$~vs.~$f$ one estimates the importance of the dipolar interaction in that sample. 
Then, using a data collapse one can estimate $\tau_0$ and obtain the mean-field interaction constants $\gamma$ defined in Eq.~(\ref{gamma}) and the mean number of nearest-neighbors $n_1$ using Eq.~(\ref{tau_DBF}). 
Finally, by fitting the collapsed data using Eq.~(\ref{4}) one is able to probe the complexity of the dipolar landscape and estimate the anisotropy constant distribution in the sample. 
We hope  that the theoretical development described here may be of use to researchers working with magnetic nanoparticles, specially in biological applications such as magnetic hyperthermia.  
Moreover, we believe that these results illustrate the enormous importance that the dipolar interaction has in most nanoparticle samples.

\begin{acknowledgements}
For their financial support, the authors would like to acknowledge the Brazilian funding agencies FAPESP,  CNPq and FAPEG,   the funding agencies Spanish MINECO and FEDER, under project MAT2014-53961-R. 
I. Andreu thanks the Spanish CSIC for her JAE Predoc contract. Finally, we thank Prof. S. M. Carneiro for her help with the TEM images of the spherical magnetite nanoparticles. 

\end{acknowledgements}

\appendix

\section{\label{app:lin}Linear Response Theory}

In this appendix we give a general derivation of Eqs.~(\ref{CR}) and (\ref{CI}), which is valid also for other materials, such as electric dipoles for instance.
Suppose a sample is subject to a time-varying field $H(t)$ pointing in a certain direction and let $\mu(t)$ denote the response of the system in this particular direction. 
Linear response theory is based on the physically reasonable assumption that  if the stimulus $H(t)$ is sufficiently small, the response $\mu(t)$ will be linear in it. 
However, this does not mean that $\mu(t)$ will depend on the instantaneous value of $H(t)$. 
It may very well depend on its past values. 
Thus, we may write in general that 
\begin{equation}\label{LRT}
\mu(t) = \int\limits_{-\infty}^t  f(t-s) H(s) \ud s =\int\limits_0^\infty  f(t') H(t-t') \ud t'
\end{equation}
where in the last equality we simply changed variables to $t' = t-s$.
The weight function $f(t-s)$ is called the  response kernel of the sample and, as far as linear response is concerned, it contains all the relevant information.

We will now analyze a series of experiments.
Suppose first that $H(t)$ was turned on at a constant value $H_0$ in the infinite past. 
So the system has had a very long time to stabilize. 
Inserting $H(t) = H_0$ in Eq.~(\ref{LRT}) then gives
\begin{equation}\label{mu_eq}
\mu_\text{eq} = \chi_0 H_0
\end{equation}
where we define the static susceptibility as 
\begin{equation}\label{chi_0_app}
\chi_0 = \int\limits_0^\infty f(t') \ud t'
\end{equation}

Next suppose that the field was turned on at a value $H_0$ for a very long time but, at $t=0$, we abruptly turn it off. 
In this case the magnetic moment will relax toward the new equilibrium value. 
This process is usually well described by the formula $\mu(t) = \mu_\text{eq} e^{-t/\tau}$, where $\tau$ is called the relaxation time of the system. 
For magnetic nanoparticles under N\'eel relaxation, it is given by Eq.~(\ref{tau1}). 
On the other hand, from Eq.~(\ref{LRT}) we now have 
\begin{equation}\label{relax}
\mu(t) = H_0 \int\limits_t^\infty f(t') \ud t'
\end{equation}
Comparing the two results we conclude that 
\begin{equation}\label{kernel}
f(t) = \frac{\chi_0}{\tau} e^{-t/\tau}
\end{equation}
The response kernel is therefore completely determined by the relaxation properties of the system.

Finally, consider an alternating magnetic field of frequency $\omega$ which has been turned on for a very long time. 
This is the situation encountered in AC susceptibility. 
Inserting $H(t) = H_0 \cos\omega t$ in Eq.~(\ref{LRT}) we recover Eq.~(\ref{mu_t}) with 
\[
\chi' = \int\limits_0^\infty \cos\omega t' f(t') \ud t' \qquad \chi'' = \int\limits_0^\infty \sin\omega t' f(t') \ud t' 
\]
Using Eq.~(\ref{kernel}) to perform the integrals, we then finally arrive at Eqs.~(\ref{CR}) and (\ref{CI}).

\bibliography{/Users/gtlandi/Documents/library}

\end{document}